\documentclass[a4paper,fleqn,usenatbib]{mnras}
\usepackage[T1]{fontenc}
\usepackage{ae,aecompl}
\usepackage{graphicx}   
\usepackage{amsmath}    
\usepackage{amssymb}    
\usepackage{enumerate}

\title[Open cluster radii from {\it Gaia} proper motions]{
A catalogue of open cluster radii determined from
\textbf{\textit{Gaia}} proper motions}

\author[N. S\'anchez, E. J. Alfaro and F. L\'opez-Mart\'inez]{
N\'estor S\'anchez$^{1}$\thanks{E-mail: nestor.sanchezd@campusviu.es},
Emilio J. Alfaro$^{2}$ and F\'atima L\'opez-Mart\'inez$^{3}$\\
$^{1}$Universidad Internacional de Valencia (VIU),
C/Pintor Sorolla 21, 46002, Valencia, Spain.\\
$^{2}$Instituto de Astrof\'isica de Andaluc\'ia, CSIC,
Glorieta de la Astronom\'ia S/N, Granada, 18008, Spain.\\
$^{3}$Centro de Estudios de F\'isica del Cosmos de Arag\'on (CEFCA),
Unidad Asociada al CSIC, Plaza de San Juan 1, 44001, Teruel, Spain.}
\date{Accepted XXX. Received YYY; in original form ZZZ}
\pubyear{2020}

\begin{document}
\label{firstpage}
\pagerange{\pageref{firstpage}--\pageref{lastpage}}
\maketitle

\begin{abstract}
In this work we improve a previously published method to
calculate in a reliable way the radius of an open cluster.
The method is based on the behaviour of stars in the proper
motion space as the sampling changes in the position space.
Here we describe the new version of the method and show its
performance and robustness. Additionally, we apply it to
a large number of open clusters using data from {\it Gaia}
DR2 to generate a catalogue of $401$ clusters with reliable
radius estimations. The range of obtained apparent radii goes
from $R_c=1.4 \pm 0.1$~arcmin (for the cluster FSR~1651)
to $R_c=25.5 \pm 3.5$~arcmin (for NGC~2437).
Cluster linear sizes follow very closely a lognormal
distribution with a mean characteristic radius of
$R_c = 3.7$~pc, and its high radius tail can be
fitted by a power law as $N \propto R_c^{-3.11\pm0.35}$.
Additionally, we find that number of members, cluster
radius and age follow the relationship
$N_c \propto R_c^{1.2\pm0.1} \cdot T_c^{-1.9\pm0.4}$
where the younger and more extensive the cluster, the
more members it presents.
The proposed method is not sensitive to low density or
irregular spatial distributions of stars and, therefore,
is a good alternative or complementary procedure to
calculate open cluster radii not having previous
information on star memberships.
\end{abstract}



\section{Introduction}

\defcitealias{Dia02}{D02}
\defcitealias{Kha13}{K13}
\defcitealias{Sam17}{S17}
\defcitealias{Can18}{C18}
\defcitealias{San18}{Paper~I}

It is well known the importance that open clusters (OCs) have 
in several areas of Astronomy, including the structure and 
evolution of the Galactic disk and the star formation process
(see, for example, reviews by \citealt{Fri95}; \citealt{Ran13}
and \citealt{Kru19}). In order to achieve more significant 
advances in these research areas, it is necessary not 
only to increase the census of known OCs but also to 
improve the determinations of their properties, such 
as distance, age, size, number of members, proper motion, 
radial velocity and reddening. The large amount of photometric 
and astrometric data publicly available online, as well as the 
current computational capabilities, have allowed the creation 
of large databases and catalogues listing the existing clusters
and their fundamental properties. Two notable examples of the 
pre-{\it Gaia} era are the widely-used catalogues published by 
\citet[][hereinafter \citetalias{Dia02}; see also
\citealt{Dia14,Sam17,Dia18}]{Dia02} and \citet[][hereinafter
\citetalias{Kha13}; see also \citealt{Kha12}]{Kha13}.
Another recent catalogue compiling
positions and multiple names for
star clusters and candidates is the one published by \citet{Bic19}.
However, it has to be mentioned that cluster properties reported
in these and other catalogues frequently differ each other
and, more importantly, the use of different data sources 
and/or methods of analysis can lead to some biases in the 
inferred cluster parameters \citep{Net15,San18,Bos19}.

The advent of the ESA's {\it Gaia} space mission \citep{Gai16} 
has opened a new era in the study of OCs. The second data release
of {\it Gaia} (DR2) \citep{Gai18} is a homogeneous source of data
with unprecedented astrometric precision and accuracy for 1.3 
billion objects. 
One of the notable outcomes of {\it Gaia}
DR2 was its immediate impact on the cluster census.
\citet{Sim19} reported more than 200 new OCs that were identified 
by simple visual inspection of the multidimensional {\it Gaia}
data (positions, proper motions and parallaxes).
\citet[][henceforth \citetalias{Can18}]{Can18}
applied the unsupervised membership assignment code UPMASK 
\citep{Kro14} to a list of 3328 known OCs and candidates
(including those contained in \citetalias{Dia02} and
\citetalias{Kha13}) and made a serendipitous discovery 
of 60 new clusters in the studied fields, whereas \citet{Liu19}
used a friend-of-friend based method to explicitly search for new 
OCs and found 76 highly probable candidates. In a recent work, 
\citet{Cas20} applied a machine learning based methodology to 
carry out a blind search for OCs in the Galactic disk. They 
first used the algorithm DBSCAN \citep{Est96} to search for 
overdensities in the five-dimensional parameter space (positions, 
proper motions, parallaxes) and then used an artificial neural 
network to confirm the cluster nature by recognizing patterns 
in their colour-magnitude diagrams. With this technique 
\citet{Cas20} reported 582 new OCs distributed 
along the Galactic disk. \citet{Can19} and \citet{Cas19} 
searched for and detected new stellar clusters towards the 
Galactic anticentre and the Perseus arm and from their results 
they concluded that the current list of known nearby OCs
is far from being complete. Since the release of 
{\it Gaia} DR2, the increase in the number of known OCs has been 
accompanied by the confirmation of non-existence of many clusters 
previously catalogued as such \citep[see for example][]{Kos18,Can20}. 
In fact, astrometric precision of {\it Gaia} DR2 has led \citet{Can20} to 
classify as not true clusters (asterisms) about a third of OCs 
listed in catalogues within the nearest 2 kpc.

Such a complex scenario (new OCs being continuously discovered 
while others being categorized as asterisms) arises together with 
the systematic and usually automated or semi-automated determination 
of OC physical properties. Nowadays, there are a variety of techniques 
and available tools that are being used to assign memberships and to 
derive OC properties as, for instance, those formerly designed by
\citeauthor{Cab85} (\citeyear{Cab85,Cab90}), UPMASK \citep{Kro14},
ASteCA \citep{Per15}, $N$-D geometry \citep{Sam16} and more
recently Clusterix \citep{Bal20}. Thanks to the quality
of {\it Gaia} DR2 data, star cluster parameters that are
being derived by the authors are the most precise to date (see 
already mentioned references). There is, however, a need for some 
caution when performing massive data processing because, as mentioned
above, slight variations in the developed strategies can lead to biases 
in the inferred cluster parameters \citep{Net15,San18}. 
Among all OC parameters that can be derived, radius is
particularly relevant.
Reliable estimates of cluster radii and member stars for a
representative sample of clusters in the Milky Way would 
allow to better identify observational constraints on the
physical mechanisms driving molecular cloud fragmentation,
the star formation process and the destruction and
dissipation of OC into the surrounding star field
\citep{Sch07,San09,Cam09,Gie18,Het19}. Additionally,
as discussed in detail in \citet{San10}, the relation
between cluster radius ($R_c$, understood in its simplest 
geometric definition as the radius of the smallest circle 
containing all the cluster stars) and the sampling radius 
($R_s$, the radius of the circular area around the cluster
position used to extract the data from the catalogue)
determines the quality of the final derived results.
The main reason for this is that
a proper estimate of OC properties
generally needs
a reliable identification of cluster members and,
depending on the method, membership assignment may be 
seriously affected if the sampling radius is either far
below (subsampled cluster) or far above (excess of field star
contamination) the actual cluster radius \citep{Sam16}.
Then, the optimal sampling radius
for studying an OC is the, in principle unknown,
cluster radius itself \citep{San10,San18}.

In order to overcome this issue we have been working
on an alternative method for inferring the radius of
an OC in an objective way without previous information
about the cluster, except for the fact that the cluster 
does exist, meaning that it is visible as an overdensity
in the proper motion space.
In this work we improve the method originally
proposed in \citet[][hereinafter \citetalias{San18}]{San18}
and apply it to the sample of OCs listed in \citetalias{Dia02}
using data from {\it Gaia} DR2.
Section~\ref{sec_metodo} describes the modified method, which
is applied in Section~\ref{sec_aplicacion} to obtain a catalogue
of OC radii.
Section~\ref{sec_analisis} is devoted to compare our results
with other catalogues whereas Section~\ref{sec_linearsizes}
analizes the obtained linear sizes and the relationship among
different cluster variables.
Finally, in Section~\ref{sec_conclusion}
we summarize our main results.


\section{Method: open cluster radii from stellar proper motions}
\label{sec_metodo}

In a first version of the method \citepalias{San18}, we
defined a transition parameter that measures the sharpness of 
cluster-field boundary in the proper motion space, and $R_c$ 
was obtained as the $R_s$ value for which the best 
cluster-field separation was achieved. The method 
was tested and applied to a sample of five OCs using 
positions and proper motions from the UCAC4 catalogue \citep{Zac13} 
and, in general, the method worked reasonably well. However, the 
strategy used in \citetalias{San18} had two limitations. Firstly,
the parameter quantifying the cluster-field transition exhibited 
significant fluctuations, making it difficult in some cases to 
identify the correct solution. With the arrival of {\it Gaia}
DR2 we realized that part of the problem was the relatively poor 
astrometric data quality, because the method was adapted and 
tested with UCAC4 proper motions, but another part of the problem 
was the definition of the transition parameter itself which had 
some sensitivity to free parameter or data variations. Secondly, 
the developed algorithm needed relatively long computation time 
to yield a valid solution because it constructed the Minimum 
Spanning Tree of each cluster several times and this is 
computationally expensive. These drawbacks made the algorithm 
unsuitable to be applied massively to OCs with data from {\it Gaia}.
For these reasons we decided to optimize the algorithm in terms 
of (a) improving its robustness to free parameters or data 
variations and (b) speeding up its execution time. Both 
requirements have been fulfilled by simplifying and optimizing 
calculations while retaining the essence of the method,
as explained below.

The general strategy
is the same: to vary
$R_s$ in a wide enough
range to be sure of including the actual cluster radius, $R_c$,
and see what happens in the proper motion space where the cluster
should be seen.
For each $R_s$ value 
there are two main steps:
(1) searching for the region covered by
the overdensity in the proper motion space and,
(2) calculating the changes in star density in this
region and in its neighbourhood as $R_s$ increases.

\subsection{Finding out the overdensity in proper motions}

In order to find out the overdensity we derive radial density
profiles for the stars in the proper motion space. If a given
starting point (star) is located in or close to the overdensity
centre then the radial profile will show an initial steep decline
followed by a shallower decrease in the region outside the 
overdensity. On the contrary, if the starting star is far
from the overdensity centre or even outside the overdensity
region then the initial decline will be less pronounced and/or
there will be irregular variations (ups and downs).
Radial density profiles are derived for all the available
stars in the proper motion space, i.e. assuming each star
as the centre of the overdensity. In each case, an
overdensity ``edge" is also determined. This edge
corresponds to the radial distance from the starting
point at which the change from an inner steep slope
to an outer shallow slope is maximum. This edge is 
meaningless if the starting point is far from the 
actual overdensity centre (irregular profiles), but
this is not important because at the end we identify 
the cluster overdensity as the one having the highest
density contrast between the overdense region and the 
background (edge), and irregular profiles will show
low contrasts. It is worth to point out that the exact
size of the overdensity (i.e. the location of its
boundary) is not needed in our method because the
condition for determining the cluster radius does
not depend critically on this choice
(see Appendix~\ref{sec_parametros}).

These calculations are performed each time the sampling
radius is increased. That is, we search for the overdensity
in the proper motion space independently for each $R_s$ and
we require that overdensity centroid remains nearly constant
for a solution to be considered valid
(see Section~\ref{sec_wellbehaved}).
To increase computational speed when calculating
density profiles we assume circular symmetry for the 
overdensity and we use concentric circular rings with the condition 
that the minimum number of stars in each ring is $N_{min}$. We keep 
$N_{min}$ as a free parameter although we have used $N_{min}=100$ 
for the final results (see Section~\ref{sec_parametros}). We have 
made several simulations by mixing different types of cluster and
field proper motion distributions and the overdensity was properly
found as long as the cluster star distribution in the proper motion
space was several times smaller than the field distribution. For
realistic gaussian distributions the algorithm finds the cluster
edge at $\sim 3$ times the cluster standard deviation. For extreme
cases, such as clusters located very close to the outermost region
of the full distribution of stars and/or samples with too low number
of stars, the overdensity edge is always found at $\sim 2.5-3.5$ 
times the cluster standard deviation. In all the tests made with
real star clusters, positions and sizes of their overdensities in
proper motion space were confirmed by eye. 

\subsection{Calculating changes in star densities}

Let us assume we have found the overdensity centroid and its (circular)
area in the proper motion space. The {\it local field} is defined 
as the concentric circular ring surrounding the overdensity and 
containing at least $N_{min}$ stars. Let us also assume that the 
sampling radius in the position space is increased by $\delta R_s$ 
arcmin and therefore the total number of stars in the proper motion
space is also increased. The question, which our method is based on,
is: how much the density of the overdensity ($D_{od}$, in stars per 
(mas/yr)$^2$) changes compared to the local field density $D_{lf}$?
If the sampling radius $R_s$ is smaller than the actual cluster 
radius ($R_c$) then $D_{od}$ will increase more than $D_{lf}$ does 
because, apart from field stars, new cluster stars are included when 
increasing $R_s$. On the other hand, if $R_s \geq R_c$, only field 
stars are included and then both $D_{od}$ and $D_{lf}$ increase by 
nearly the same amount. This last assertion is true as long as the 
region covered by the overdensity and the local field is relatively 
small in comparison with the total sample distribution, that is, as 
long as the local average density variation is not significant. 
Field density gradients did not affect the method performance 
because local densities are always estimated on relatively small 
regions and averaging over the densest (toward the field distribution 
peak) and less dense (toward the opposite direction) parts.

In order to properly deal with uncertainties we assume Poisson 
statistic when calculating overdensity and local field densities. 
However, apart from possible statistical fluctuations, the local 
field may exhibit density variations along the ring surrounding 
the overdensity due to variations in the underlying field 
distribution. This effect may be relevant if, for instance, 
the overdensity is located very close to the outermost part 
of the star field distribution or at any region with a 
relatively high field density gradient. In order to take 
this into account, we calculate many times the local field 
density on different random ring quadrants and we consider 
the uncertainty associated with the field density to be the 
maximum and minimum obtained values along the ring.

\subsection{Workflow}

Omitting minor details of the algorithm, we span a wide
range of $R_s$ values and, at each step, search for the 
overdensity and calculate both overdensity and local field 
density changes ($\Delta D_{od}$ and $\Delta D_{lf}$, 
respectively).
The general workflow can be summarized as follows:
\begin{enumerate}[(1)]
\item An initial $R_s$ values is set and proper
motions are read for all stars corresponding to that sampling.
\item Starting on each of the stars, radial
density profiles in the proper motion space are derived,
including their centres and edges.
\item The best overdensity is selected as that
exhibiting the highest average-to-edge density contrast.
\item Density changes for this overdensity
($\Delta D_{od}$) and its local field ($\Delta D_{lf}$)
are calculated.
\item Set $R_s=R_s + \delta R_s$ and go back
to step (1).
\end{enumerate}
Finally, the results are processed and
the cluster radius $R_c$ is assigned as the $R_s$ value
from which $\Delta D_{od} \simeq \Delta D_{lf}$.
Taking into account the associated
uncertainties, we actually report lower and upper
limits for fulfiling this condition
(see Section \ref{sec_aplicacion}).

With the changes implemented we were able to improve the method 
presented in \citetalias{San18}, making it more robust against
variations of the free parameters (see Section~\ref{sec_parametros}).
Moreover, by eliminating the use of Minimum Spanning Trees, we also
sped up the algorithm and the execution is now around 17 times faster
than the previous version making feasible its application to large 
databases, which is the main goal of this work.

\section{Application to clusters with proper motions from
\textbf{\textit{Gaia}}}
\label{sec_aplicacion}

We applied the proposed method to all OC listed in the 
\citetalias{Dia02}'s catalogue. The current version of
this catalogue (V3.5) available through
VizieR\footnote{http://vizier.u-strasbg.fr} \citep{Och00}
contains updated information on 2167\footnote{It
should be pointed out that entries 1016
and 1017 in this catalogue correspond to the
same object (FSR~1496) and that some objects are duplicates
under different names \citep{Bic19}.}
optically visible OCs and candidates, including a 
compilation of their angular apparent diameters.
Using the cluster coordinates and a maximum sampling radius 
of four times the radius reported in \citetalias{Dia02},
we extracted positions and proper motions of all sources
from the {\it Gaia} DR2 catalogue.
We did not apply any magnitude cut or filtering
in proper motion error of the {\it Gaia} DR2 data.
Then we executed our algorithm over all
the clusters with $R_s$ spanning across
all their possible values. A total of 401 OCs yielded valid 
solutions in this first massive application of our method.
In this section we first show some examples of
different kinds of obtained solutions, and then we present 
the final cluster radii catalogue (\ref{sec_catalog}).

\subsection{Well-behaved solutions}
\label{sec_wellbehaved}

\begin{figure*}
\includegraphics[width=\columnwidth]{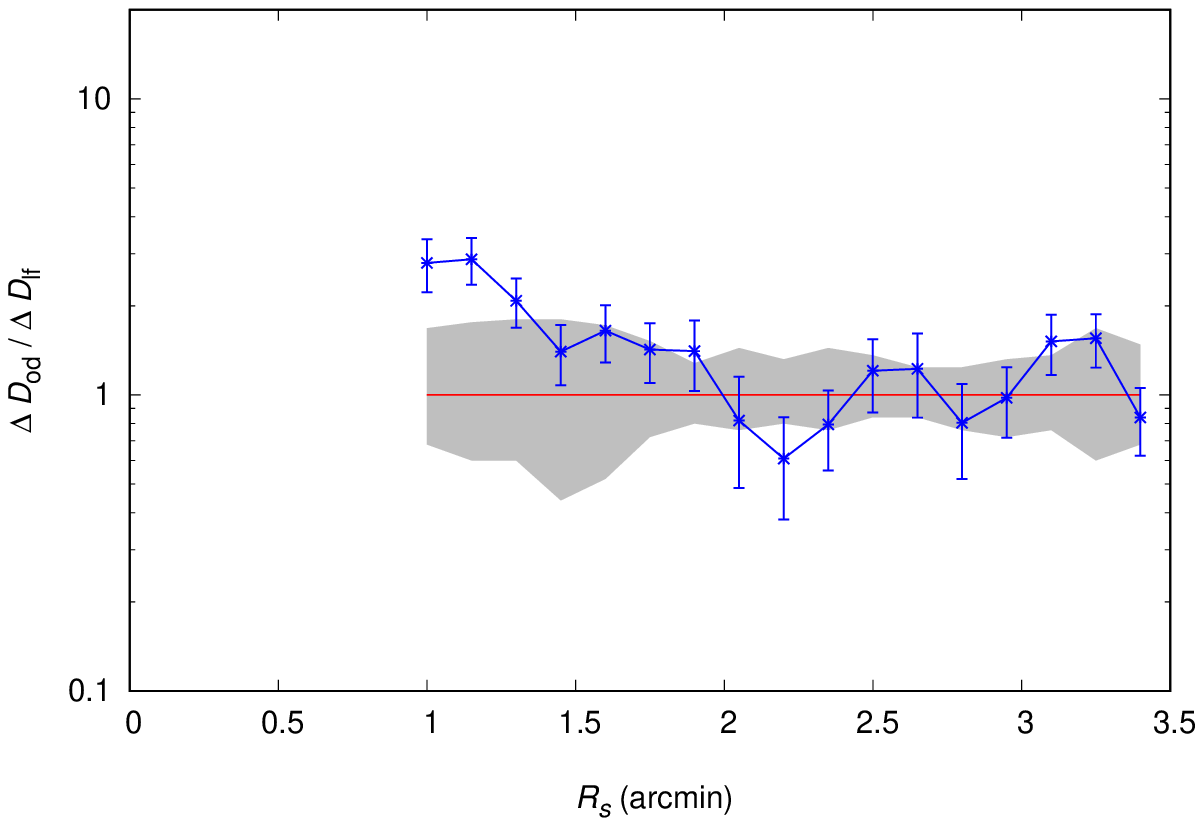}
\includegraphics[width=\columnwidth]{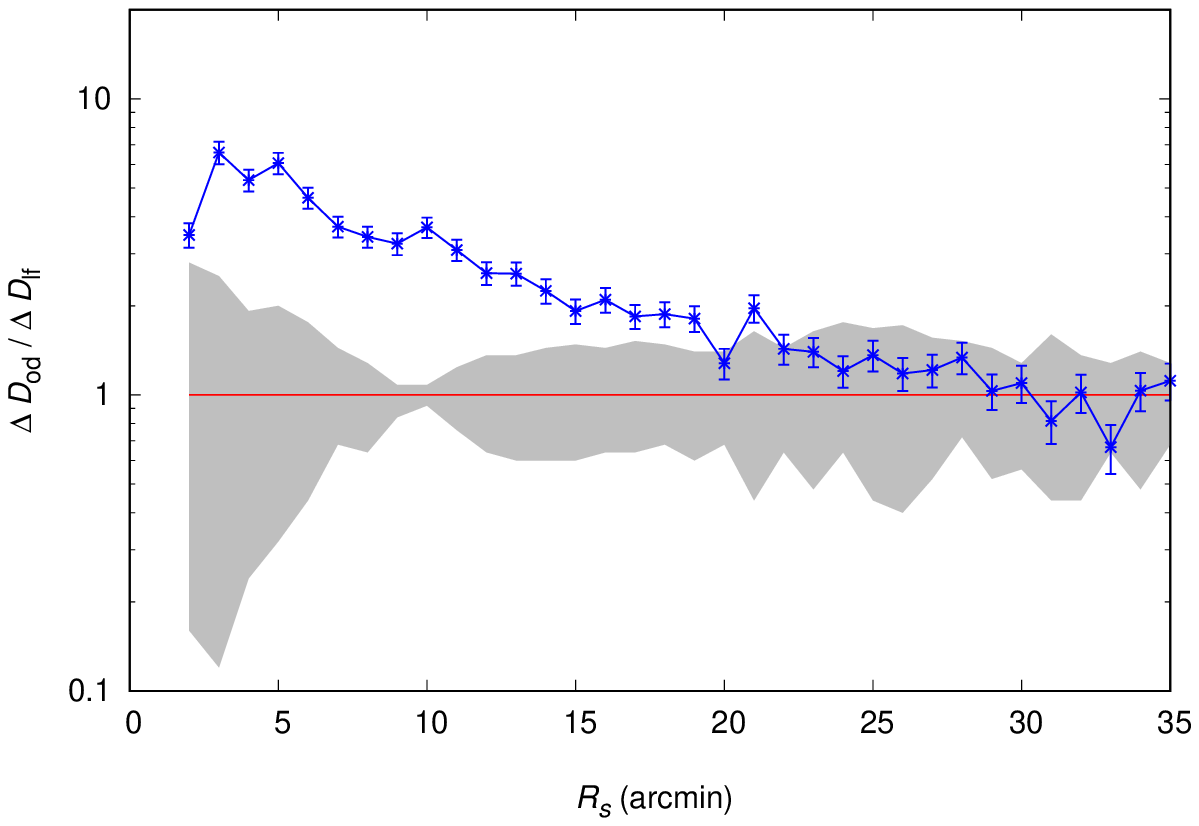}
\includegraphics[width=\columnwidth]{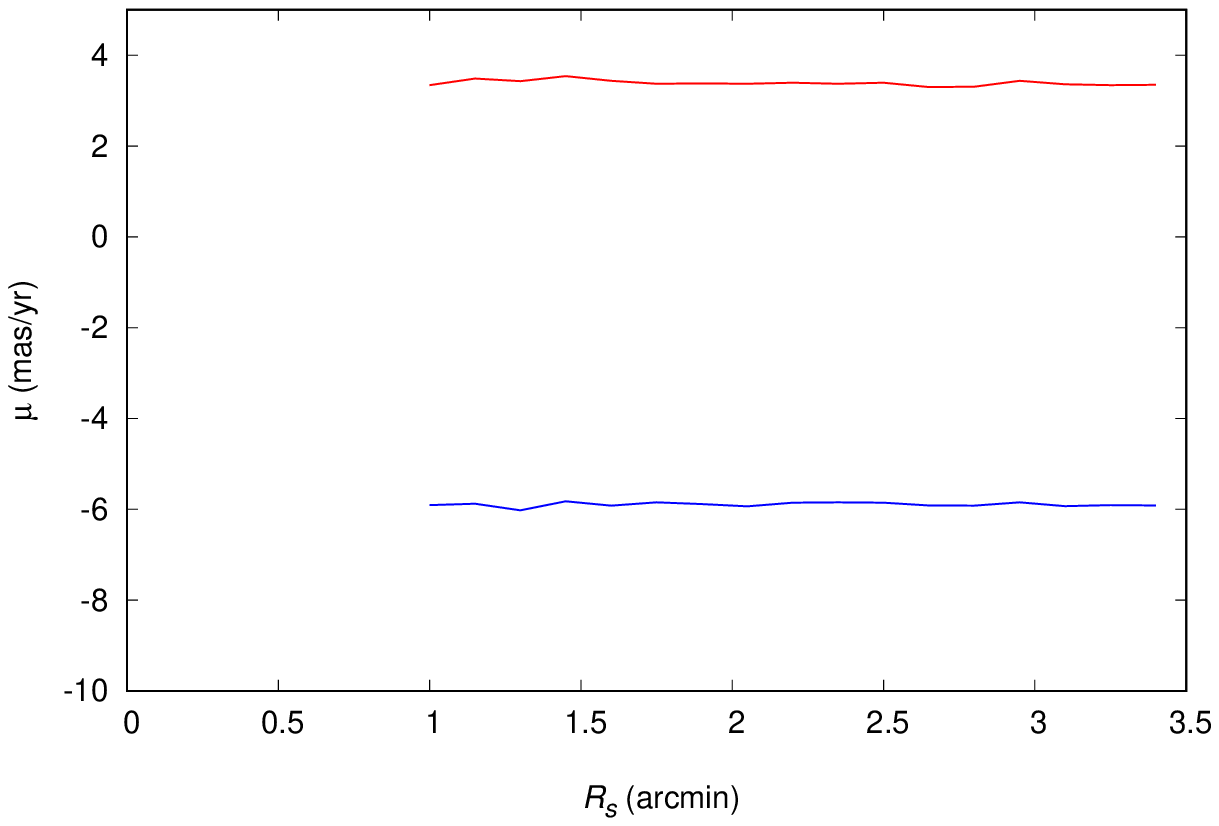}
\includegraphics[width=\columnwidth]{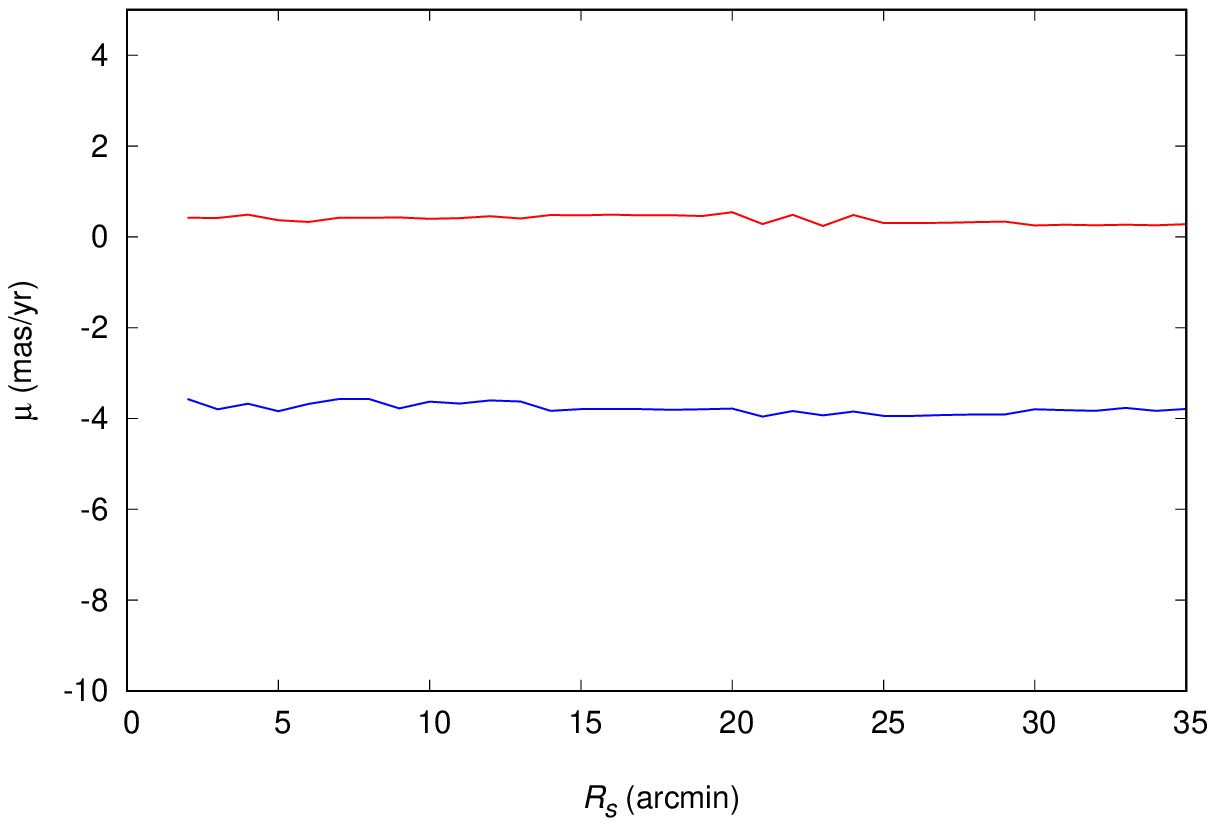}
\caption{Results of applying the proposed method to the open 
cluster NGC~3255 (left-hand panel) and NGC~2437 (right-hand 
panel). Upper panels: blue symbols represent the ratio between 
density variation in the overdensity region ($\Delta D_{od}$) 
and density variation in the local field region ($\Delta D_{lf}$) 
for different sampling radius values ($R_s$). Error bars are 
estimated by assuming Poisson statistics. Red line, shown as 
reference, corresponds to the case $\Delta D_{od}=\Delta D_{lf}$ 
expected when $R_s$ reaches the actual cluster radius $R_c$. Grey 
area indicates the uncertainty associated to the local field 
(details in the text). Lower panels: the corresponding overdensity 
proper motion centroid in right ascension (blue line) and 
declination (red line) as a function of $R_s$.}
\label{fig_eta_A}
\end{figure*}

Figure~\ref{fig_eta_A} displays two clusters for which the method 
found valid and ``well behaved" solutions (we refer these as 
type A solutions).
These two examples correspond to the type A results having the 
smallest (NGC~3255) and the highest (NGC~2437) found cluster 
radii. In order to be considered type A, a solution should 
fulfil two conditions: density changes in the overdensity 
region should decrease gradually from $\Delta D_{od} > 
\Delta D_{lf}$ to $\Delta D_{od} \simeq \Delta D_{lf}$, as 
we can see in upper panels of Figure~\ref{fig_eta_A} and, 
additionally, overdensity centroid in the proper motion 
space should be unequivocally determined (lower panels). 
For the open cluster NGC~3255, the first valid sampling 
occurs at $R_s=1$~arcmin and in this case we get $\Delta 
D_{od}/\Delta D_{lf} \simeq 3$. This means that, as the 
sampling radius increases, the density of the overdensity 
increases around three times faster than the local field 
density. This is because, besides field stars, new cluster 
stars are being included as $R_s$ increases and, therefore, 
we are still in the $R_s < R_c$ region in the position space. 
In spite of fluctuations, the expected general trend toward similar 
density change values is clearly observed for NGC~3255. Around 
$R_s=1.3$~arcmin blue symbols go into the grey region representing 
the local field uncertainty and around $R_s=2.1$~arcmin they reach 
the red line corresponding to the $\Delta D_{od} = \Delta D_{lf}$ 
case.
We reflect
these uncertainties in the final 
cluster radius estimation. In the case of NGC~3255, we get 
$R_c = 1.3-2.1$~arcmin which is above the value of $1$~arcmin 
indicated in \citetalias{Dia02}, around the $1.5$~arcmin estimated 
by \citet[][hereinafter \citetalias{Sam17}]{Sam17}
and clearly below the $\sim 8$~arcmin
reported by \citetalias{Kha13}. The highest obtained 
$R_c$ value was for the open cluster NGC~2437 (right panel in 
Figure~\ref{fig_eta_A}). The execution of the algorithm for this 
better sampled cluster clearly starts in the region $R_s < R_c$ 
with $\Delta D_{od} \simeq 8\Delta D_{lf}$ and, always with the 
already mentioned criteria for the lower and upper limits, it 
returns the solution $R_c = 22.0-29.0$~arcmin. This range of 
values is higher than values reported by \citetalias{Dia02} 
($10$~arcmin) and \citetalias{Sam17} ($17$~arcmin) for this OC
but smaller than the one given by \citetalias{Kha13} 
($34$~arcmin).

\begin{figure*}
\includegraphics[width=\columnwidth]{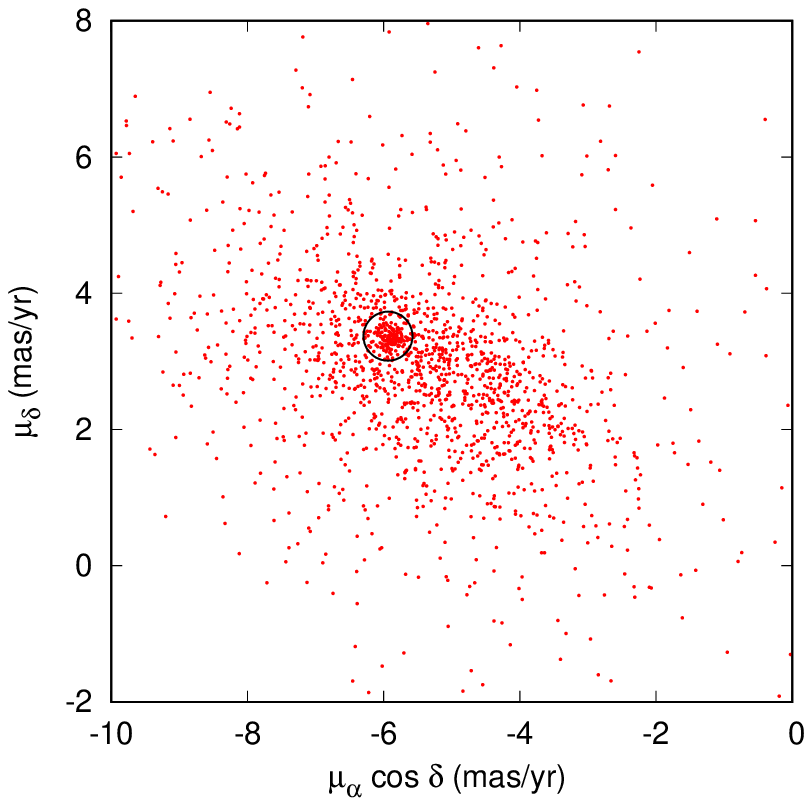}
\includegraphics[width=\columnwidth]{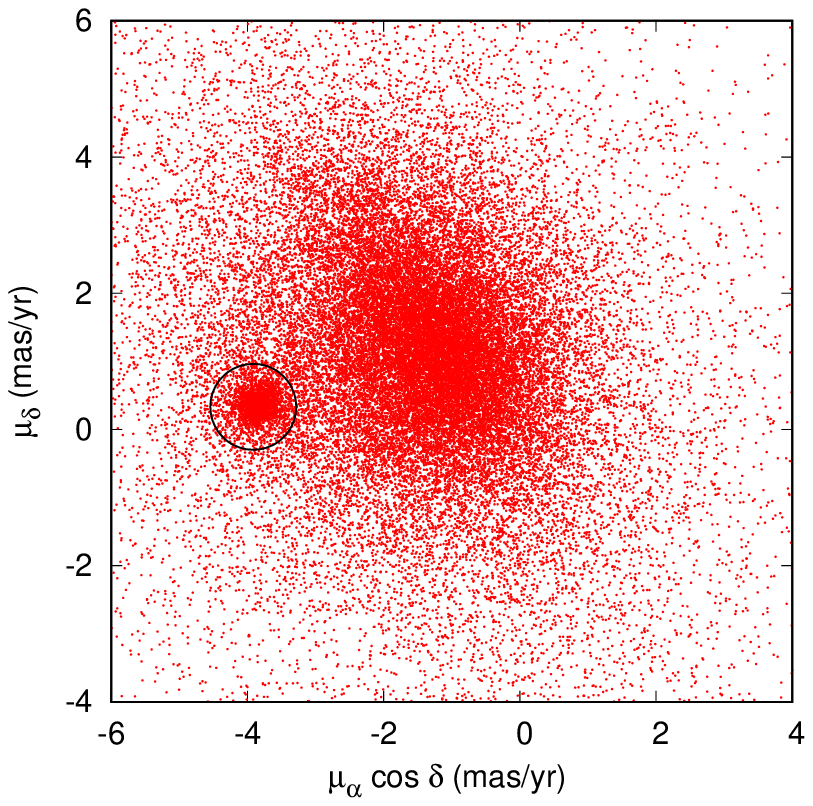}
\includegraphics[width=\columnwidth]{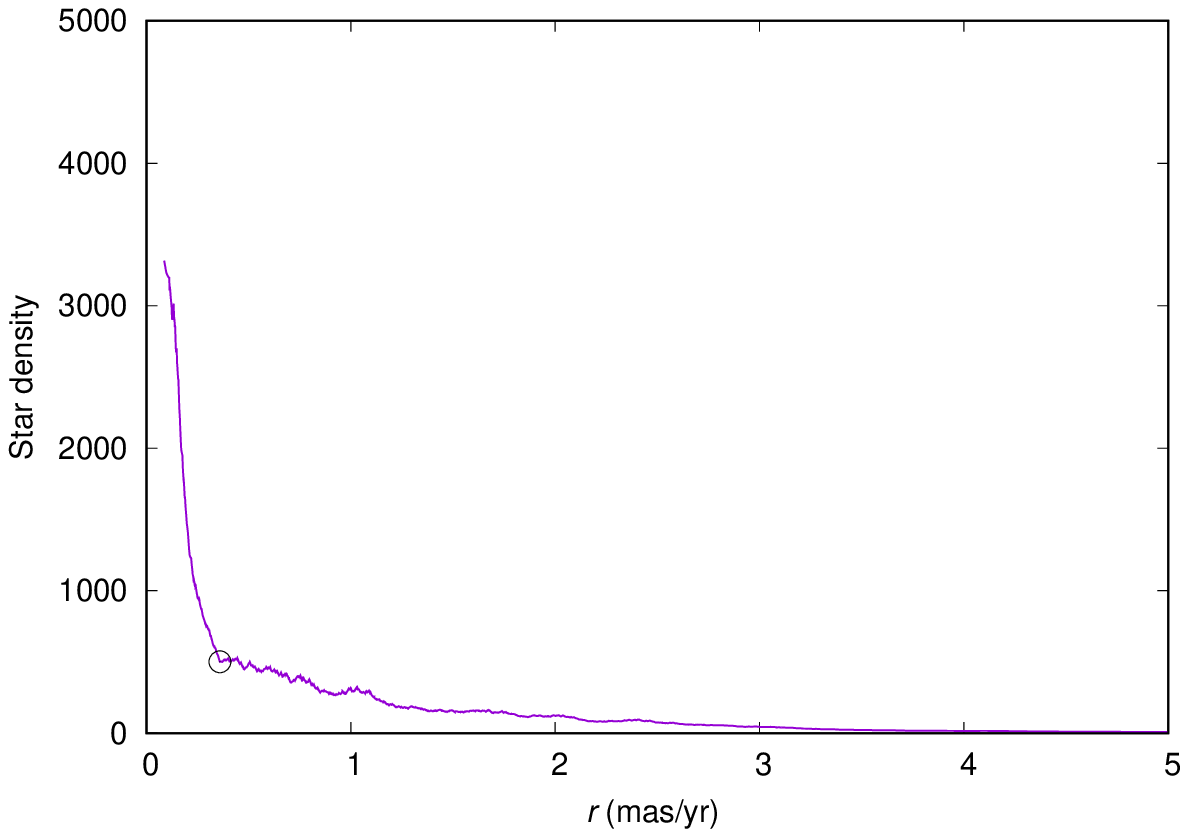}
\includegraphics[width=\columnwidth]{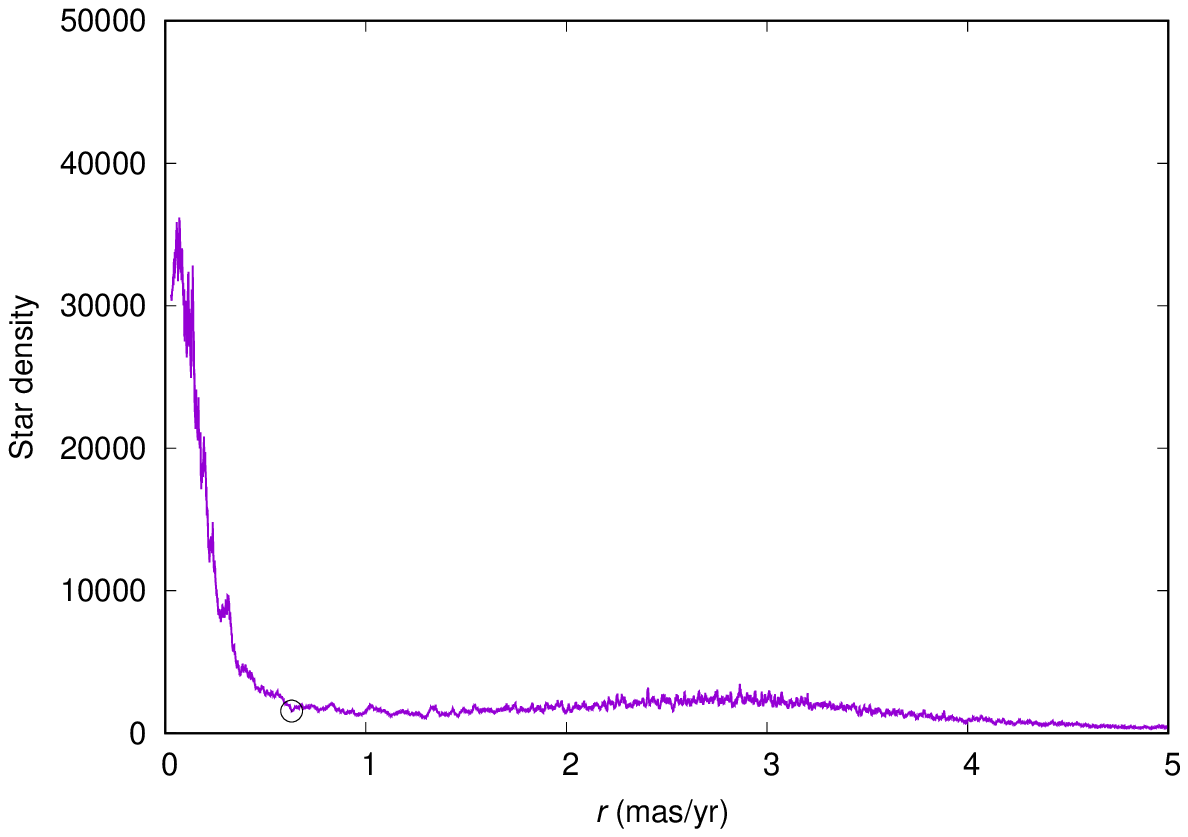}
\caption{Upper panels: distribution of proper motions for stars
in the open cluster NGC~3255 (left panel) y NGC~2437 (right panel) 
when the sampling radius equals the upper limit of the determined 
radius: $R_s=2.1$ and $R_s=29.0$, respectively. Black circles 
indicate positions and radii of the overdensity determined by 
the algorithm. Lower panels: the corresponding radial density 
profiles in stars/(mas/yr)$^2$ for the overdensities used to 
find out their positions and edges (indicated as little open 
circles over the profiles).}
\label{fig_vpd}
\end{figure*}

For both clusters shown in Figure~\ref{fig_eta_A},
centroids are found always at the same position
(see lower panels) which is one of the conditions to 
be fulfilled in order to be considered a valid solution.
Centroid of NGC~3255 is at $(\mu_{\alpha}\cos\delta,\mu_{\delta})
=(-5.93,+3.37)$ mas~yr$^{-1}$ whereas for NGC~2437 is at
$(\mu_{\alpha}\cos\delta,\mu_{\delta})= (-3.91,+0.33)$
mas~yr$^{-1}$. These centroids has been properly found as
can be seen in Figures~\ref{fig_vpd}, where star proper
motion distributions are shown for these two clusters.
Overdensity ``edges" can be clearly seen in the radial density 
profiles (lower panels in Figures~\ref{fig_vpd}) and they are 
marked with little open circles on the profiles. These points 
are used as references to estimate overdensity (inside the edges)
and local field (outside but close to the edges) densities and
their changes at each iteration.

\subsection{More uncertain solutions}

\begin{figure*}
\includegraphics[width=\columnwidth]{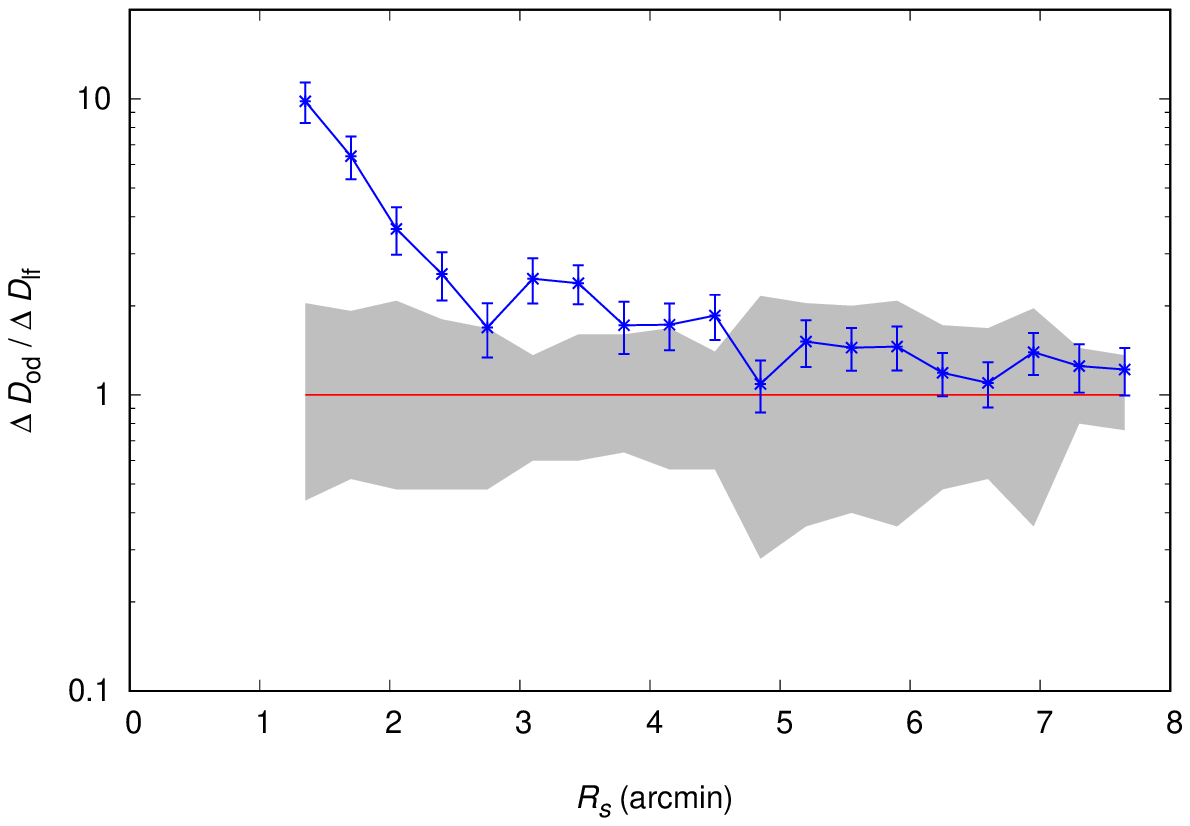}
\includegraphics[width=\columnwidth]{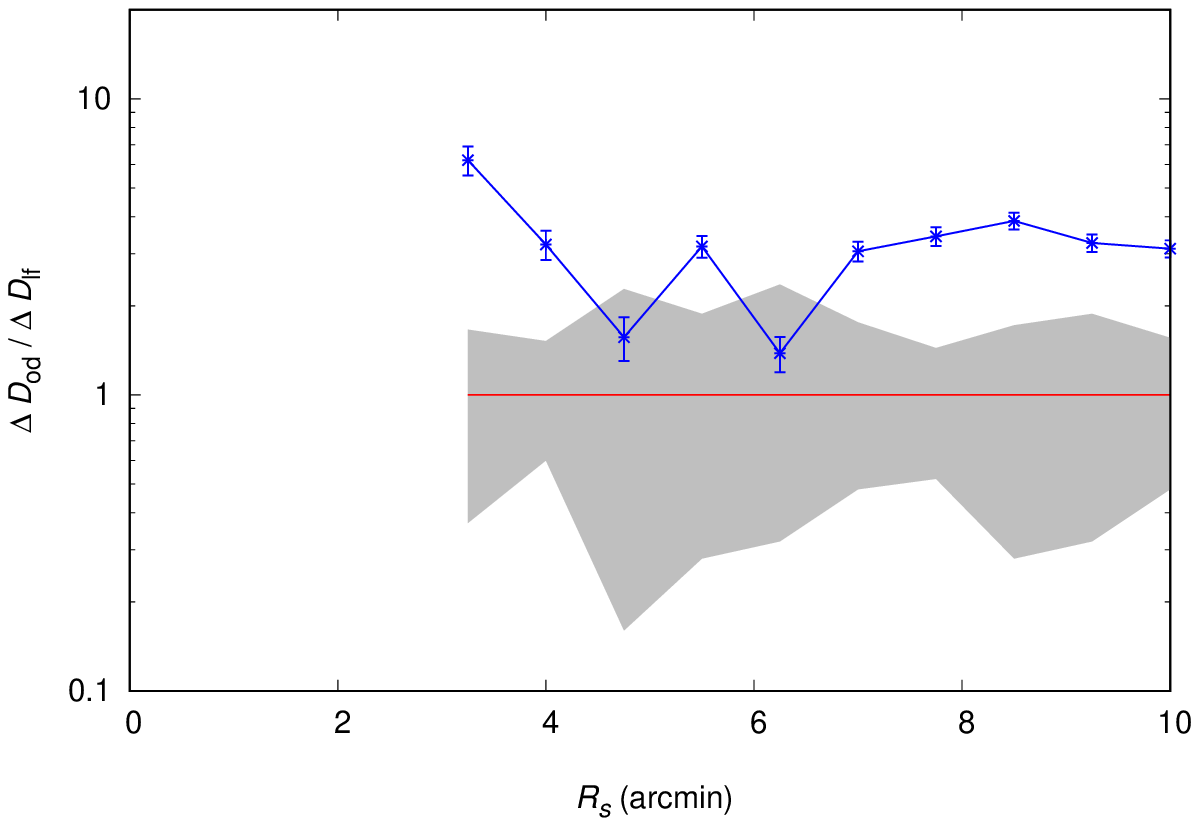}
\includegraphics[width=\columnwidth]{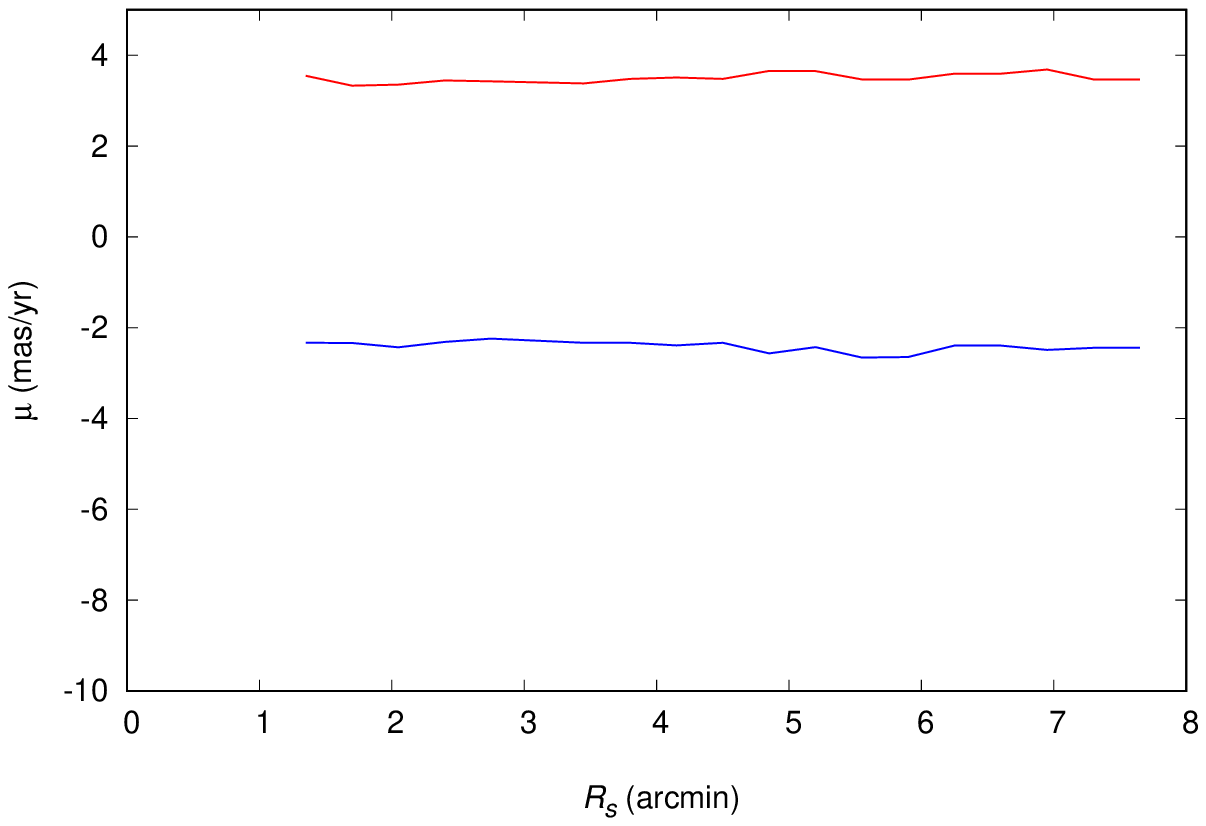}
\includegraphics[width=\columnwidth]{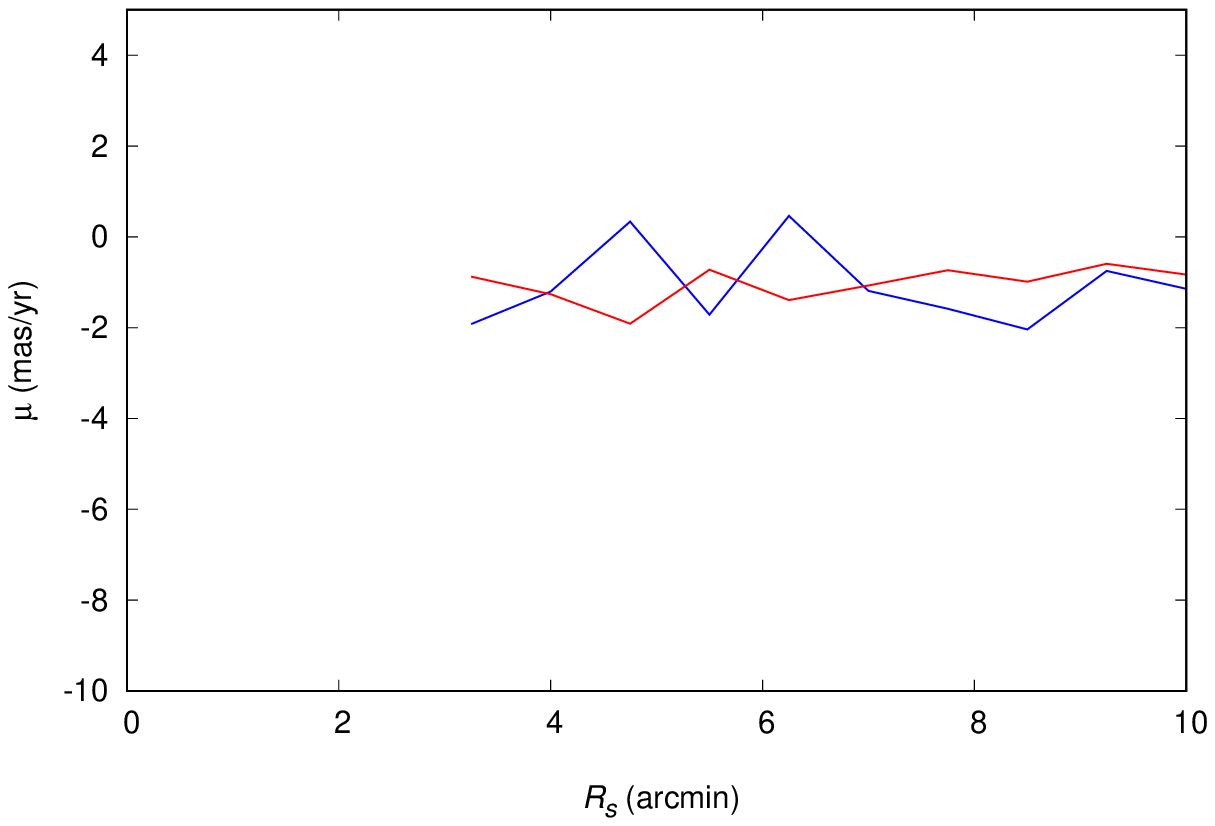}
\caption{As in Figure~\ref{fig_eta_A} but for open cluster
NGC~2453 (left panels) and NGC~7801 (right panels).}
\label{fig_eta_BC}
\end{figure*}

Some obtained results are not so well behaved as those shown in 
Figure~\ref{fig_eta_A}. An example can be seen in left panels of 
Figure~\ref{fig_eta_BC}, corresponding to the open cluster NGC~2453.
Even though the algorithm found a suitable and robust cluster 
centroid in the proper motion space (lower left panel),
$\Delta D_{od} / \Delta D_{lf}$ exhibits noticeable 
fluctuations that make it difficult to clearly constrain lower and 
upper $R_c$ limits. This kind of solutions has been flagged as type 
B, meaning that we found a valid solution but that $R_c$ estimation 
is more uncertain than in type A solutions. In order to deal 
with these fluctuations, but maintaining objective criteria, we 
demand at least two consecutive intersections with grey area or 
red line for estimating lower and upper radius limits. In any 
case, all our results were checked by eye to identify type A and 
B solutions and to confirm that lower and upper radius limits 
really represent the change from decreasing to nearly constant 
behaviour in the $\Delta D_{od} / \Delta D_{lf}$--$R_s$ plot. 
With these criteria, we get $R_c = 4.95 \pm 1.35$~arcmin for 
NGC~2453 (upper right panel in Figure~\ref{fig_eta_BC}).
Despite the associated uncertainty, this value is certainly
higher than the $2.0-2.5$~arcmin indicated by \citetalias{Dia02}
and \citetalias{Sam17} but in agreement with the $4.8$~arcmin 
assigned by \citetalias{Kha13}. A more recent work \citep{Gon19} 
based on {\it Gaia} DR2 suggests a higher value, in the range
$8-10.5$~arcmin.

\subsection{Undetected solutions}

\begin{figure}
\includegraphics[width=\columnwidth]{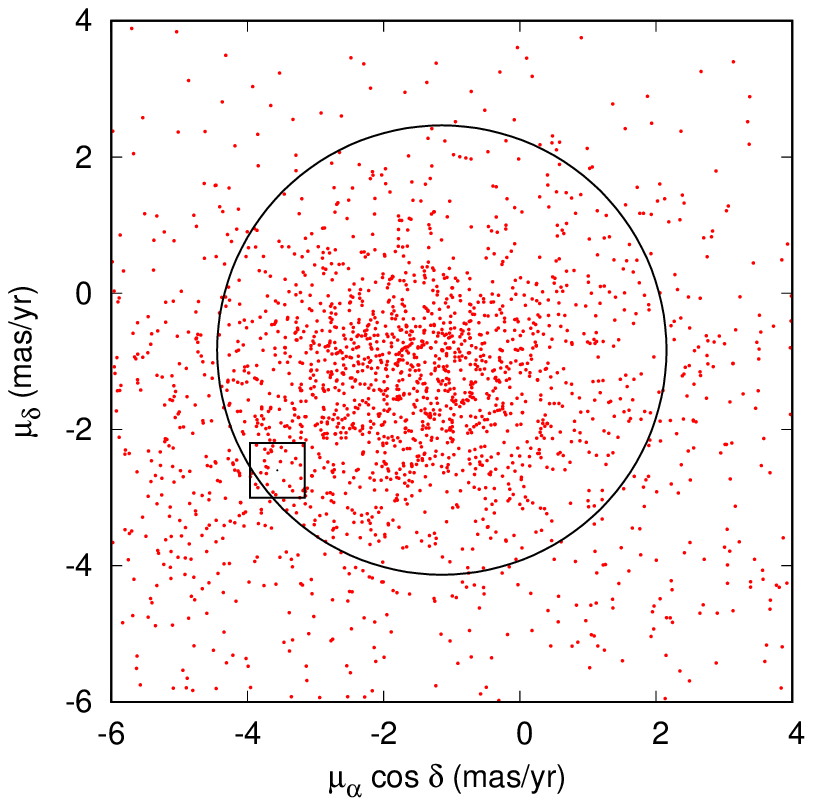}
\caption{Proper motion distribution of stars in the region of 
NGC~7801 using a sampling radius of $10$~arcmin. Black circle 
indicates position and size of the overdensity calculated by 
the algorithm, that obviously corresponds to the maximum of 
the field star distribution. Black square indicates, as 
reference, centroid position according to the \citetalias{Dia02}'s 
catalogue.}
\label{fig_vpd_7801}
\end{figure}

We are not reporting results of cluster for which the algorithm
did not converge to a valid solution. These cases require 
further analysis in order to verify whether additional data 
processing can ensure convergence or, by the contrary, whether 
there is a physical cause for the non-convergence (for instance 
a complex proper motion structure or that there is no open cluster 
at all). Right panels in Figure~\ref{fig_eta_BC} show the first 
entry for which we did not find a valid solution corresponding 
to the open cluster NGC~7801. According to \citetalias{Dia02} and 
\citetalias{Sam17} its radius is $4.0-4.5$~arcmin whereas for 
\citetalias{Kha13} it is $9.4$~arcmin. The no-solution is seen 
in the facts that the result $\Delta D_{od} \simeq \Delta 
D_{lf}$ is never reached and that the overdensity centroid 
is not properly found (it is fluctuating in the range $\sim 
0-2$~mas/yr). Proper motions of stars in the region of NGC~7801 
using $R_s=10$~arcmin are shown in Figure~\ref{fig_vpd_7801}.
No overdensity is visible by eye in the proper motion space, 
apart from the maximum of the full distribution, and therefore 
the algorithm is no able to find a valid solution.
The reason is that NGC~7801 is an asterism,
as originally suggested by \citet{Sul73} and recently
confirmed by \citet{Can20}.

There may be different reasons for not
finding a valid solution.
Firstly, 
relatively low spatial star densities and/or small angular cluster 
sizes translate into a first valid $R_s$ value above the actual 
cluster radius. Secondly, there are some cluster catalogued in
\citetalias{Dia02} that does not show (by eye) any clear 
overdensity in the proper motion space with data from {\it Gaia}
DR2. These last cases should be analyzed separately in future 
studies in order to ascertain whether they are star clusters or
asterisms \citep[see for example asterisms reported by][]{Can20}.
There is a third kind of non-valid solution having a recognizable 
overdensity in proper motion space but that, for some (other) 
reason, do not reach the condition $\Delta D_{od} \simeq \Delta 
D_{lf}$ and that will be addressed 
in future works.


\subsection{Catalog of open cluster radii}
\label{sec_catalog}

The application of the proposed method to the sample
of OCs listed in \citetalias{Dia02} has allowed us to build
a catalogue with 401 reliable radius values determined in a 
systematic and independent way through the star proper
motions.
Main outputs of the algorithm are lower ($R_{low}$) and upper 
($R_{opt}$) limits of the cluster radius estimation. According 
to discussion in \citet{San10} and \citetalias{San18}, the upper
limit $R_{opt}$ would be the optimal sampling radius needed to
be sure of including all cluster stars but minimizing the number
of field stars contaminants. In our final catalogue we indicate 
the estimated cluster radius as a central value
$R_c = (R_{opt} + R_{low})/2$ and an associated 
uncertainty $\Delta R_c = (R_{opt} - R_{low})/2$.
We also report the cluster proper motion 
$(\mu_{\alpha}\cos\delta,\mu_{\delta})$ estimated for the optimal 
case $R_s = R_{opt}$. Additionally, using the area covered by the 
overdensity ($A_{od}$), its mean density ($D_{od}$) and also 
the local field density ($D_{lf}$), we can make an estimation of 
the number of kinematic cluster member: $N_c = (D_{od} - D_{lf}) 
A_{od}$, which will be an upper limit of the actual number of 
members because additional criteria (for instance parallaxes or 
photometry) may exclude some stars and because the actual cluster 
radius may be smaller than $R_{opt}$.
The final results are shown in Table~\ref{tabcumulos} and include 
OC name, equatorial coordinates (J2000), obtained cluster radius 
($R_c$) and its associated uncertainty ($R_{err}$), a flag indicating 
type of solution (A or B), the estimated number of kinematic member 
stars and the mean cluster proper motion.
\begin{table*}
\centering
\caption{Catalog of cluster radii. This is only a portion for
guidance regarding its form and content. The full table is
available online.}
\label{tabcumulos}
\begin{tabular}{lcccccccc}
\hline
Name & RA    & DEC   & $R_c$    & $R_{err}$ & Type    & $N_c$ &
     $\mu_\alpha\cos\delta$ & $\mu_\delta$ \\
     & (deg) & (deg) & (arcmin) & (arcmin)  &         &       &
     (mas~yr$^{-1}$) & (mas~yr$^{-1}$) \\
\hline
Berkeley~58  & 0.05000 & 60.96667 & 10.50 & 2.50 & B & 357 & -3.387 & -1.820 \\
Berkeley~59  & 0.55833 & 67.41667 &  6.70 & 1.90 & A & 297 & -1.608 & -2.040 \\
Berkeley~104 & 0.87500 & 63.58333 &  2.75 & 0.25 & A &  57 & -2.449 & +0.129 \\
Berkeley~1   & 2.40000 & 60.47500 &  3.05 & 0.25 & A &  76 & -2.726 & -0.101 \\
King~13      & 2.52500 & 61.16667 &  6.65 & 2.05 & B & 534 & -2.815 & -0.794 \\
Berkeley~60  & 4.42500 & 60.93333 &  3.25 & 0.25 & A & 116 & -0.629 & -0.682 \\
FSR~0486     & 5.08750 & 59.31806 &  2.70 & 0.50 & A & 106 & +0.119 & -0.056 \\
Mayer~1      & 5.47500 & 61.75000 &  4.90 & 3.30 & B &  91 & -3.213 & -1.482 \\
SAI~4        & 5.91667 & 62.70389 &  2.45 & 0.35 & B & 319 & -2.492 & -0.608 \\
Stock~20     & 6.31250 & 62.61667 &  3.60 & 0.90 & A &  79 & -3.319 & -1.235 \\
\hline
\end{tabular}
\end{table*}

\section{Comparison with other catalogues}
\label{sec_analisis}

\subsection{Angular radii}

The range of obtained radii goes from $R_c=1.4 \pm 0.1$~arcmin for 
the open cluster FSR~1651 (with only $N_c=118$ kinematic members) 
to $R_c=25.5 \pm 3.5$~arcmin for NGC~2437 ($N_c= 1891$).
It is not a straightforward task to compare our values with
those obtained in other studies because the concept of radius
is ambiguous itself (it depends on the cluster morphology and
structure) and its definition often differs among authors.
As mentioned before, we are using the simple, geometric
approach in which cluster radius is defined as the radius
of the smallest circle containing all assigned members,
what we called {\it covering} radius in \citetalias{San18}.
Other characteristic radii are the core radius,
half-mass (or half-light) radius, tidal radius
and the commonly used radial density profile radius,
defined as the radius where the cluster surface density
drops below field density. Mixing different concepts can 
lead to inaccurate or biased analysis \citep[see discussions
in][]{Mad12,Pfa16}. For example, if most of the OCs follow
smooth radial density profiles with very low projected
densities in the outer parts, then it is possible that
radius values determined from these profiles are
systematically below real extents of the clusters
(i.e. covering radii as determined here).

The last homogeneous derivation of memberships and OC
properties using data from {\it Gaia} DR2 was made by
\citetalias{Can18}. They used proper motions and parallaxes
to identify members and, from there, to derive very
precise properties for a total of 1229 clusters.
However, they did not report cluster radii. Their
radius $r50$, that containing 50\% of the members, is
not a reliable description of the total cluster
extent. In fact, the first systematic cluster size
determination based on {\it Gaia} DR2 is presented
in this work.
We then compare our results with radius values
from \citetalias{Dia02}, \citetalias{Kha13} and
\citetalias{Sam17}, which were estimated in
different ways. Radii in \citetalias{Dia02} are
just a bibliographic data compilation and, as such,
they are heterogeneous with respect to the methods 
used for estimating them, that include visual inspection. 
\citetalias{Kha13} used spatial, kinematic and photometric 
data from PPMXL \citep{Roe10} and 2MASS \citep{Skr06} to
assign memberships and then fitted King's \citep{Kin62} profiles 
to determine cluster radii in a uniform and homogeneous way.
From the fitting, \citetalias{Kha13} obtained
the radius for the core (r0), for the central part (r1) and
for the cluster (r2). We use the last one for the comparison.
On the other hand, \citet[][hereinafter \citetalias{Sam17}]{Sam17} 
used star positions given in UCAC4 catalogue \citep{Zac13}
to estimate cluster angular radii through a careful visual 
inspection of radial density profiles. 

\begin{figure*}
\includegraphics[width=\columnwidth]{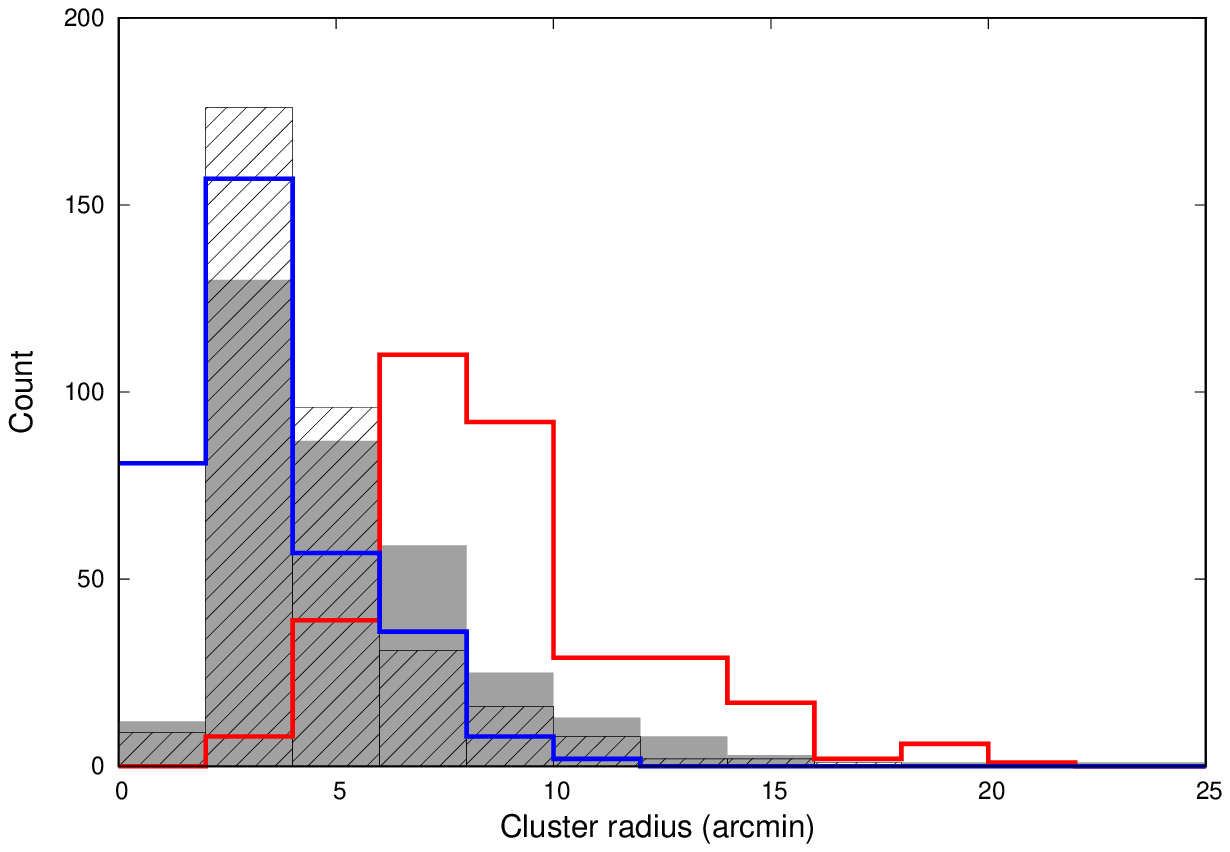}
\includegraphics[width=\columnwidth]{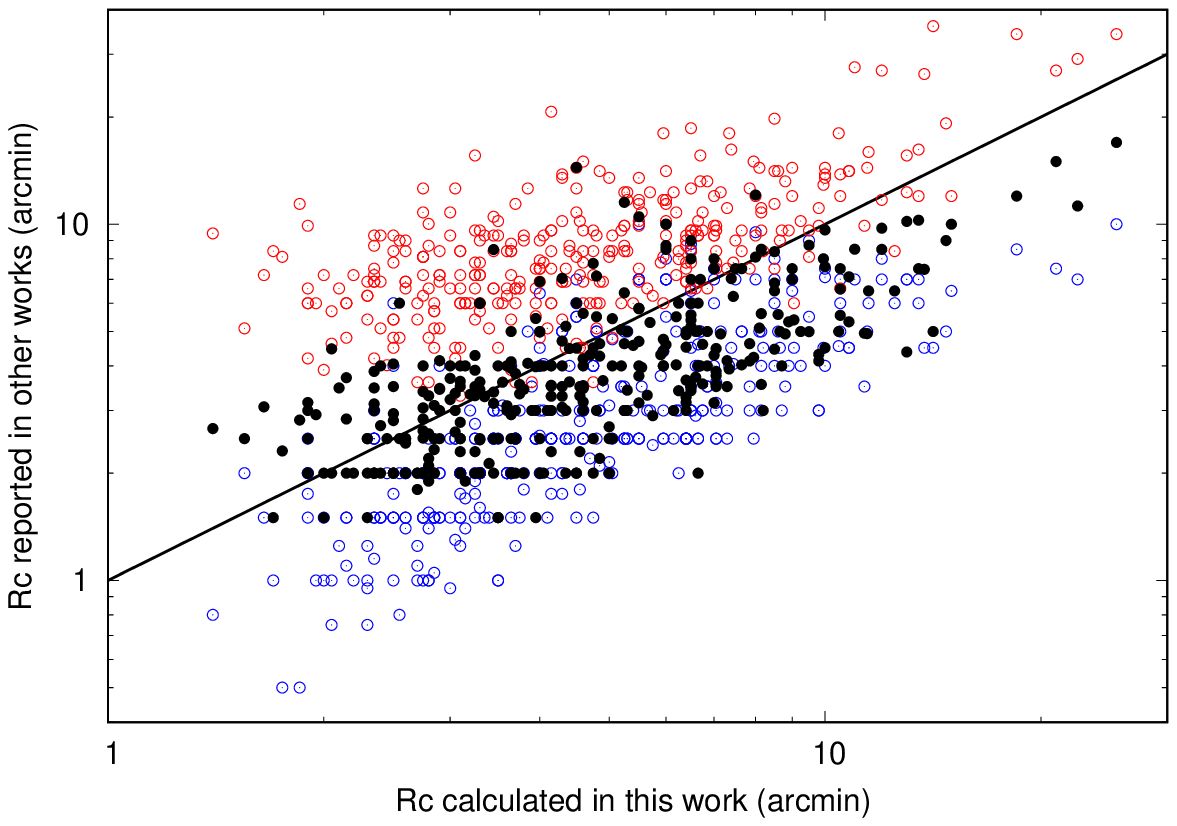}
\caption{Left panel: Distributions of apparent angular
cluster radii estimated in different works: this work
(grey bars), \citetalias{Dia02} (blue line), \citetalias{Kha13}
(red line) and \citetalias{Sam17} (bars with a diagonal line
pattern). Right panel: cluster radii reported in
\citetalias{Dia02} (blue open circles), \citetalias{Kha13}
(red open circles) and \citetalias{Sam17} (black solid
circles) as function of the value estimated in
this work. Solid line indicates the 1:1 relation.}
\label{fig_histo}
\end{figure*}

In order to compare our results with \citetalias{Dia02},
\citetalias{Kha13} and \citetalias{Sam17}, we crossmatched
the full lists of objects
among these catalogues. From the 401 cluster with valid
solutions (also included in \citetalias{Dia02}), there are
341 that have radius values reported both in \citetalias{Kha13}
and in \citetalias{Sam17}.
Figure~\ref{fig_histo} compares cluster radius values
for these 341 common OCs.
We can see clear offsets, being \citetalias{Dia02}'s
values consistently smaller and \citetalias{Kha13}'s values
consistently larger than our results.
Interestingly, there seems to be a good agreement, with
no apparent bias, between cluster radii estimated by 
\citetalias{Sam17} and our results, each of which is
based on different methods and datasets. 

\subsection{Background-corrected radial density profile}

\begin{figure}
\includegraphics[width=\columnwidth]{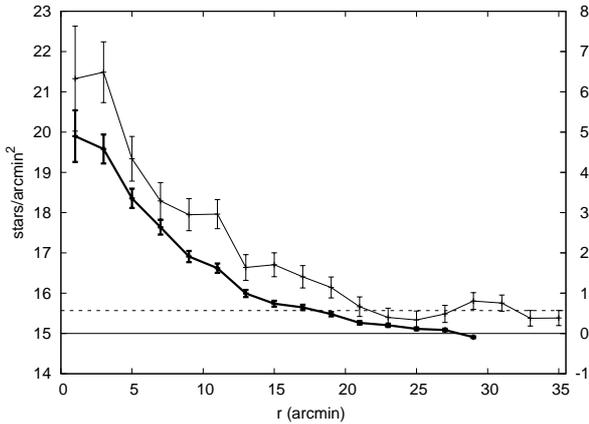}
\caption{Spatial density profile of stars toward NGC~2437
obtained by using data from the {\it Gaia} DR2.
Thin black line (corresponding to left y-axis) shows
the resulting profile for the full sample of stars and
the horizontal dashed line indicates the mean value of
$\sim 15.7$~stars/arcmin$^2$ calculated in the outermost
ring ($30-35$~arcmin).
Black thick line (right y-axis using the same units) is an
estimation of the actual (clean) cluster density profile by
using only stars located inside the proper motion overdensity
(see upper right panel in re~\ref{fig_eta_A}) and removing
the estimated proportion of field stars. Vertical
bars represent Poissonian statistical errors.}
\label{fig_perfilXY}
\end{figure}

In general, angular sizes obtained with
the proposed method agree very well with those reported
by \citetalias{Sam17} (Figure~\ref{fig_histo}), although
many particular cases may, of course, differ.
\citetalias{Sam17}'s radii were estimated using radial
density profiles in the position space. Regardless of 
the used method, we would expect similar results for the
same clusters.
For example, for the cluster NGC~2437 (Figure~\ref{fig_eta_A})
we get $R_c=25.5 \pm 3.5$~arcmin, clearly above the value
$17$~arcmin reported by \citetalias{Sam17}.
We have calculated the
spatial radial density profile for this cluster with the
{\it Gaia} DR2 data in the usual way, that is, by counting
stars in 2-arcmin width concentric rings around the cluster 
centre. The 
profile
is shown in Figure~\ref{fig_perfilXY}.
Strictly speaking, density profile (that includes both cluster 
and field stars) merges into the background at $r=20-22$~arcmin. 
We have also estimated a background-corrected radial density 
profile. This was done by using only stars located in the proper
motion overdensity region and subtracting the proportion of field 
star contamination, which is estimated based on the local 
field-overdensity ratio of densities. The background density 
was also estimated with the corresponding proportion of field 
stars. The ``clean" profile of spatial density of member stars 
(black thick line in Figure~\ref{fig_perfilXY}) 
reaches the zero density level around $\sim 27-29$~arcmin.
NGC~2437 radius obtained through spatial density profile is in
agreement with the result yielded by our algorithm and the 
difference with \citetalias{Sam17}'s result seems to be more
related to the differences in the used datasets.

\subsection{Other outputs}

Another algorithm output is the number of estimated kinematic
members (number of overdensity stars corrected by subtracting the
expected number of field stars). \citetalias{Can18} assigned
memberships applying an unsupervised algorithm to proper motions
and parallaxes from {\it Gaia}. Their initial sampling radius
were based on \citetalias{Dia02} and \citetalias{Kha13}
catalogues but they claim that, in principle, membership
determination is little affected by the exact sampling as
long as the full cluster is sampled \citep{Kro14}.
Our number of members ($N_c$) correlate very well with
\citetalias{Can18}'s values ($N_{C18}$) in such
a way thet the best linear fit passing through
the origin is $N_c = (1.24 \pm 0.02) \cdot N_{C18}$.
This means that, on average, we are selecting $\sim$25\%
more members than \citetalias{Can18}, something that can
be explained by the fact that they included the parallax
as an additional discriminant variable
and that they restricted their study to stars brighter
than $G=18$.
Finally, the algorithm also provides cluster proper motions,
i.e. centre positions of the overdensities.
In Table~\ref{tabmovpro} we compare our results with
\citetalias{Dia02}, \citetalias{Kha13}, \citetalias{Sam17}
and \citetalias{Can18} for each of the $\sim 300$ clusters
in common with these catalogues.
\begin{table}
\centering
\caption{Differences between mean proper motions obtained in this work
(S20) and those published by \citetalias{Dia02}, \citetalias{Kha13},
\citetalias{Sam17} (with their three methods: M1, M2 and M3) and
\citetalias{Can18}. Proper motions are in mas~yr$^{-1}$ and
the number following the plus/minus symbol is one standard
deviation.}
\label{tabmovpro}
\begin{tabular}{lcc}
\hline
   & $\Delta\mu_\alpha\cos\delta$ & $\Delta\mu_\delta$ \\
\hline
S20-D02     & $-0.23 \pm  2.42$ & $+0.23 \pm  2.49$ \\
S20-K13     & $+0.21 \pm  2.62$ & $-0.02 \pm  2.78$ \\
S20-S17(M1) & $-0.57 \pm  2.17$ & $+0.20 \pm  2.46$ \\
S20-S17(M2) & $-0.44 \pm  1.95$ & $+0.28 \pm  2.04$ \\
S20-S17(M3) & $-0.35 \pm  2.30$ & $+0.54 \pm  2.31$ \\
S20-C18     & $-0.04 \pm  0.50$ & $-0.01 \pm  0.29$ \\
\hline
\end{tabular}
\end{table}
Generally speaking, our results are consistent with these
previous studies with differences, on average, smaller than
one standard deviation. As expected, the strongest agreement
is with \citetalias{Can18} who used data from {\it Gaia}.
A slight trend is noticeable in which differences with
\citetalias{Dia02} and \citetalias{Sam17} are opposite
to those with \citetalias{Kha13}. It seems that the
zero-point differences with UCAC4 and PPMXL catalogues
affect more than the different methodologies used
in these works.

At this point we have to stress that our method is
intended basically to determine OC radii, but the
comparison of other derived properties with existing 
data allow us to check the reliability of our results.

\ \\
\section{Analysis of linear sizes}
\label{sec_linearsizes}

\begin{figure}
\includegraphics[width=\columnwidth]{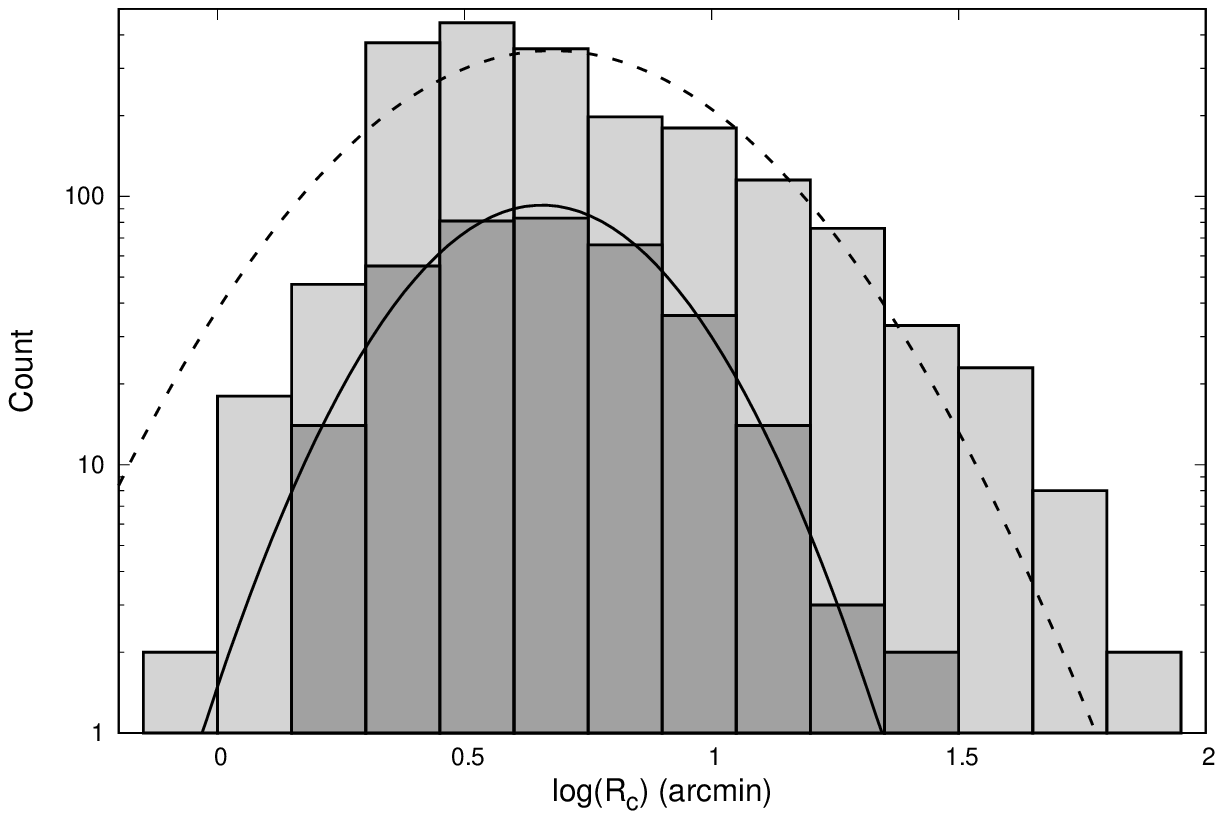}
\includegraphics[width=\columnwidth]{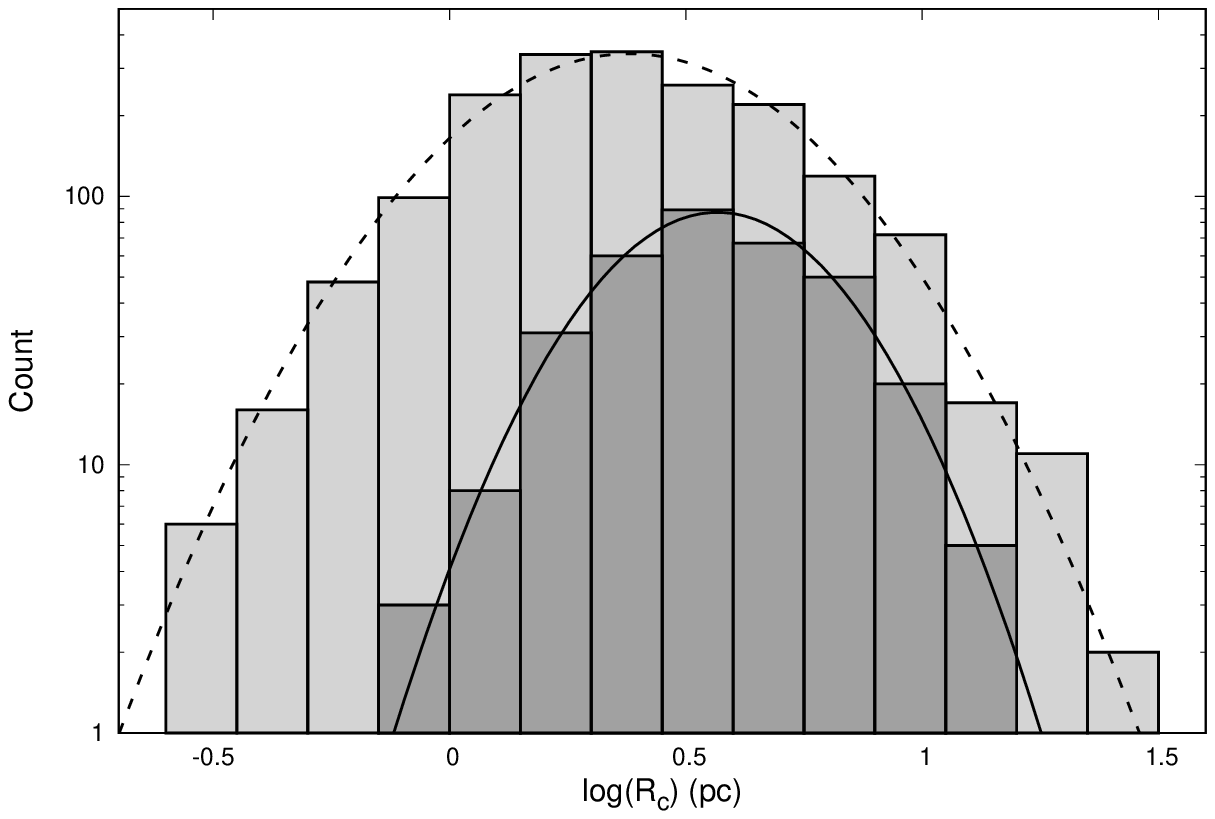}
\caption{Upper panel: angular size distributions for the
full sample of OCs in \citetalias{Sam17} (light grey bars)
and for the subsample with radius values obtained in 
this work (dark grey bars). Bin size is 0.15~dex.
Best fits to lognormal functions are shown as
dashed (\citetalias{Sam17}) and solid (this work)
lines.
Lower panel: the same distributions but for linear
radii (in pc) obtained by \citetalias{Sam17} with
distances catalogued in \citetalias{Dia02} and by us
with distances reported by \citetalias{Can18}.}
\label{fig_histoRlog}
\end{figure}

We have mentioned that angular sizes obtained in this work
agree very well with those reported by \citetalias{Sam17}
(Figure~\ref{fig_histo}), even though they were calculated
by using different procedures and datasets.
When plotted in a log-log plot (upper panel in
Figure~\ref{fig_histoRlog}) both distributions follow very
similar patterns. Mean radius for both cases is almost
the same (around $4.7-4.8$~arcmin).
To analize the linear sizes we cross-matched our results
with \citetalias{Can18}'s catalogue because they determined
very precise OC distances from {\it Gaia} DR2 parallaxes.
Lower panel in Figure~\ref{fig_histoRlog} shows the
result for the 334 clusters in common. We see that the 
distribution follows very well a lognormal function. This
is the kind of distribution that best fit star cluster 
populations in external galaxies \citep[for instance in
M~51][]{Sch07}. The mean of the distribution is
$\log(R_c) \simeq 0.57$ with a standard deviation
$\simeq 0.23$. This characteristic cluster size
of $R_c \simeq 3.7$~pc is not very different from
the value $3.94 \pm 0.12$~pc found by \citet{Lar04}
for the effective 
radius\footnote{Note that our radii
should be necessarily higher than the effective
(half-light) radii as defined by \citet{Lar04}.}
of OC systems in a sample of $15$ external galaxies
observed with the HST. For comparison, we also show
in the lower panel of Figure~\ref{fig_histoRlog} the
resulting distribution of \citetalias{Sam17} using
distances from \citetalias{Dia02}. It also follows
a well defined lognormal distribution but the mean
values differ by 0.2~dex (mean radius in
\citetalias{Sam17} is $\simeq 2.4$~pc).
If we only compare distance distributions for 321
cluster that are present in both samples we obtain a
very similar difference: our results are practically
the same and mean radius in \citetalias{Sam17} is
$\simeq 2.5$~pc.
Then, linear sizes estimated in this work are on
average 55\% higher than the ones in \citetalias{Sam17}.
The comparison between the individual distances given by
\citetalias{Dia02} and \citetalias{Can18}, for our samples,
shows very similar values (difference smaller than 7\% on
average, distances in \citetalias{Can18} are higher than
those in \citetalias{Dia02}), which is unable to generate
the displacement in linear sizes observed in
Figure~\ref{fig_histoRlog} (lower panel).
Summarizing, the application of our methodology to a
sample of clusters as those listed by \citetalias{Sam17},
but using astrometric data from {\it Gaia} DR2, generates
a subsample with precise values of angular sizes and any
apparent bias, taken individually.
However, the comparison between the distributions of
angular and linear radii of the \citetalias{Sam17} 
catalogue and ours (Figure~\ref{fig_histoRlog}) shows
that our most reliable results are obtained for clusters
with larger diameters.

\subsection{Power-law fitting}

\begin{figure}
\includegraphics[width=\columnwidth]{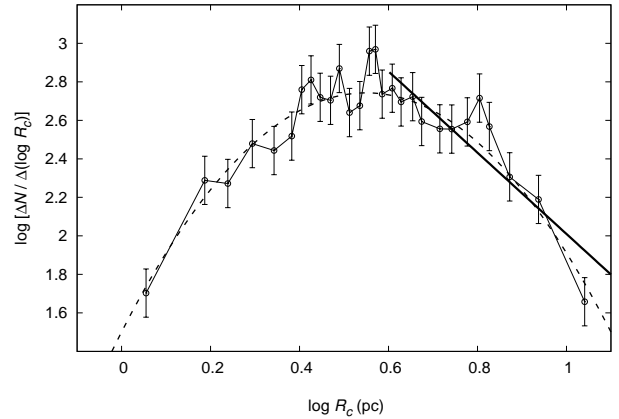}
\caption{Cluster radii distribution (number
of clusters per unit logarithmic radius interval).
The bin size is such that the number of data points
per bin is always $12$. Dashed line corresponds to the
best lognormal fitting for the full sample. Solid line
stands for the power-law fitting starting from $R_c=4$~pc,
that yields $\alpha = -3.11 \pm 0.35$.}
\label{fig_histoEAJ}
\end{figure}

Some authors claim that the high-values tail of the radius
distribution can be described by a power law of the form
$N(R_c) \sim R_c^\alpha$, being $N(R_c)$ the number of
objects with radius $R_c$ per unit linear size interval.
We may expect that young, new-born clusters roughly follow
the same distribution of giant molecular clouds, for which
on average $\alpha=-3.3\pm0.3$ \citep{Elm96,San06}.
It is possible, however, that the star formation process
itself and/or the non-uniform early evolution of OCs
drastically change or even erase this initial scenario.
\citet{Bas05} found a of $\alpha=-2.2\pm0.2$ for
stellar clusters in the disk of the galaxy M~51.
For our analysis, we have chosen to fit the number of
clusters per unit {\it logarithmic} radius interval versus
the logarithm of the radius. In such a plot, the power-law
tail would be a straightline with slope $\alpha+1$.
In addition, we do no set a constant bin size but a
constant number of clusters in each bin. It has been 
proven that this kind of variable sized binning
yields bias-free and robust estimates, especially
for small sample sizes \citep{Mai05}.
We carried out several tests varying the minimum
radius for the fitting ($3.5$ or $4.0$ pc) and the
number of points per bin (between $12$ and $16$),
and the slopes obtained were between $-2.03$ and
$-2.11$ (with errors between $0.26$ and $0.44$).
Figure~\ref{fig_histoEAJ} shows the result for
a lower limit of $4.0$~pc and $12$ data points
per bin.
This result is fully compatible with the distribution
of molecular clouds ($\alpha \sim -3$). However, it is
worth noting that a lognormal function is a much better
description for the whole cluster radii distribution
(dashe line in Figure~\ref{fig_histoEAJ}).
Trying to fit a power-law to this kind of distribution
is not a suitable approach, although in principle it
could be valid when only part of the information
(biased toward high radius values) is available.

\subsection{$N_c$-$T_c$-$R_c$ relation}

Now we proceed to examine the relationship among different
variables, in particular number of members ($N_c$), cluster
radius ($R_c$) and age ($T_c$). $N_c$ is related to the total
cluster mass although, in principle, such a connection is not
straightforward because we are dealing with a wide range of
ages and galactocentric distances, and both dynamical and
evolutionary effects may influence the stellar mass function.
Relations between the mass (or number of stars), radius and
age have been observed for young clusters 
\citep[see, for example,][]{Pfa09,Pfa16}.
Similarly, it has been suggested that Galactic OCs spanning
a variety of ages and properties exhibit the same type of
scaling relations. There seems to be some correlations
between mass, size and age, although there is still
considerable uncertainty, especially about the effect
of age on the cluster mass or size \citep[see, for 
example,][and references therein]{Sch06,Cam09,Jos16,Gun17}.
In order to perform this analysis
we have adopted cluster ages from \citet{Bos19}, who
used {\it Gaia} DR2 astrometric and photometric data to
derive precise ages for a sample of 269 OCs, from which
we have 63 clusters in common. We constructed a
multivariable
linear model incorporating the variables $\log N_c$,
$\log R_c$ (pc) and $\log T_c$ (Myr), and the best 
fit to the data yields:
\[
N_c \propto R_c^{1.2\pm0.1} \cdot T_c^{-1.9\pm0.4}
\]
On average, larger OCs have more stars, but additionally
younger clusters also tend to contain more stars, i.e. tend
to be more massive. This trend agrees with the result
obtained by
\citet{Jos16}
and, in general, with the idea
that OCs dissolve slowly with time
\citep{Wie71}.
Clearly, there are many physical processes acting
(simultaneously or at different times) and their
mixed effects may spuriously create, amplify, or
diminish this kind of relationships.


\section{Conclusions}
\label{sec_conclusion}

In this work we improve the method proposed in a previous paper
\citepalias{San18} for objectively calculating the radius of an
open cluster using star positions and proper motions. The method
spans the sampling radius around the cluster centre, identifies
the cluster overdensity in proper-motion space and compares the
changes in star densities between the overdensity and its 
neighbourhood as the sampling radius increases. The key point
of the method is the assumption that these changes should be 
similar when the sampling radius equals (or is close) to the 
actual cluster radius. Here we significantly improved the 
method making it faster than the previous version
\citepalias{San18} and much less sensitive to variations 
of free parameters.

Additionally, we applied the method to all $2167$ open clusters 
catalogued by \citetalias{Dia02}, using proper motions from the 
{\it Gaia} DR2. From this we obtained a catalogue of $401$ open 
clusters with reliable radius values calculated with the proposed
procedure. On other hand, many of the clusters that did not yield
a valid solution do not seem to show an overdensity in proper
motions when are seen with data from {\it Gaia} DR2 and their
true nature should be investigated.
The general distribution of angular radii agrees reasonably
well with that obtained by \citetalias{Sam17}, whereas some
offsets
are observed when compared with catalogues of \citetalias{Dia02}
and \citetalias{Kha13}. The obtained distribution of cluster
proper motions is consistent with those obtained by
\citetalias{Dia02}, \citetalias{Kha13} and \citetalias{Sam17},
and it is very similar to that reported in \citetalias{Can18}.
Calculated linear sizes follow a lognormal distribution with a 
mean value of $R_c = 3.7$~pc, and this distribution shows a
shift to higher values with respect to the corresponding
\citetalias{Sam17} distribution. The high radius tail of
obtained distribution can be fitted by a power law of the
form $N \propto R_c^{-3.11\pm0.35}$.
We also found that, on average, younger clusters tend
to contain more stars according to the relation
$N_c \propto R_c^{1.2\pm0.1} \cdot T_c^{-1.9\pm0.4}$,
in general agreement with some previous works.

Although the exact behaviour of the algorithm is in some way
related to cluster spatial density profile, the proposed
method is mainly focused on what happens in proper motions
rather than in spatial positions and, therefore, is not
sensitive to factors such as low spatial densities or
irregular distributions of stars. The only condition of 
the method to work properly is that the cluster must be 
visible as an overdensity in the proper motion space.
Thus, this method is a good alternative or complement
to the standard radial density profile approach.


\section*{Acknowledgements}

We are very grateful to the referee for his/her careful
reading of the manuscript and helpful comments and
suggestions, which improved this paper.
NS and EJA acknowledge support from the Spanish Government
Ministerio de Ciencia, Innovaci\'on y Universidades through
grant PGC2018-095049-B-C21 and from the State Agency for
Research of the Spanish MCIU through the ``Center of 
Excellence Severo Ochoa" award for the Instituto de
Astrof\'isica de Andaluc\'ia (SEV-2017-0709).
F.~L.-M. acknowledges partial support by the
Fondos de Inversiones de Teruel (FITE).

\appendix

\section{Effect of varying the input parameters}
\label{sec_parametros}

We have carried out several tests to evaluate the performance
of the proposed method and to verify that the algorithm does
not yield biased results and does not critically depend on
input
parameters. Tests involved both simulated and real clusters with
different characteristics including sizes, number of data points,
and overdensity position in the field distribution. In general,
the method works well on all cases as long as overdensity is 
visible\footnote{It depends on each case, but in our simulated
tests this condition usually means that star cluster density
in the proper motion space must be at least as dense as field
star density in the same position.} in the proper motion 
space. 
We have two relevant free parameters already mentioned in
Section~\ref{sec_metodo}: the minimum number of data points 
allowed to estimate the density in a given region ($N_{min}$)
and the step for spanning the sampling radius ($\delta R_s$).

\subsection{Parameter $N_{min}$}

$N_{min}$ determines, among other things, the sample (bin size)
of overdentity profiles (lower panels in Figure~\ref{fig_vpd})
making it smoother of noisier. This may affect the exact location
of the overdensity ``edge". If the overdensity edge determined
at a given step is a little closer or further than its ``real" 
position then estimations of overdensity and local field 
densities will vary. However, the core of the method is
to compare density {\it variations} as $R_s$ increases,
and the condition of similar variations
($\Delta D_{od} \simeq \Delta D_{lf}$)
when $R_s \gtrsim R_c$ will be fulfilled independently of
the exact edge position. In fact, this condition should be 
fulfilled in any two relatively small and adjacent regions as 
long as no new cluster stars are included as $R_s$ increases.
Therefore, the exact location of the edge
has no effect on the final cluster radius obtained.

\begin{figure*}
\includegraphics[width=\columnwidth]{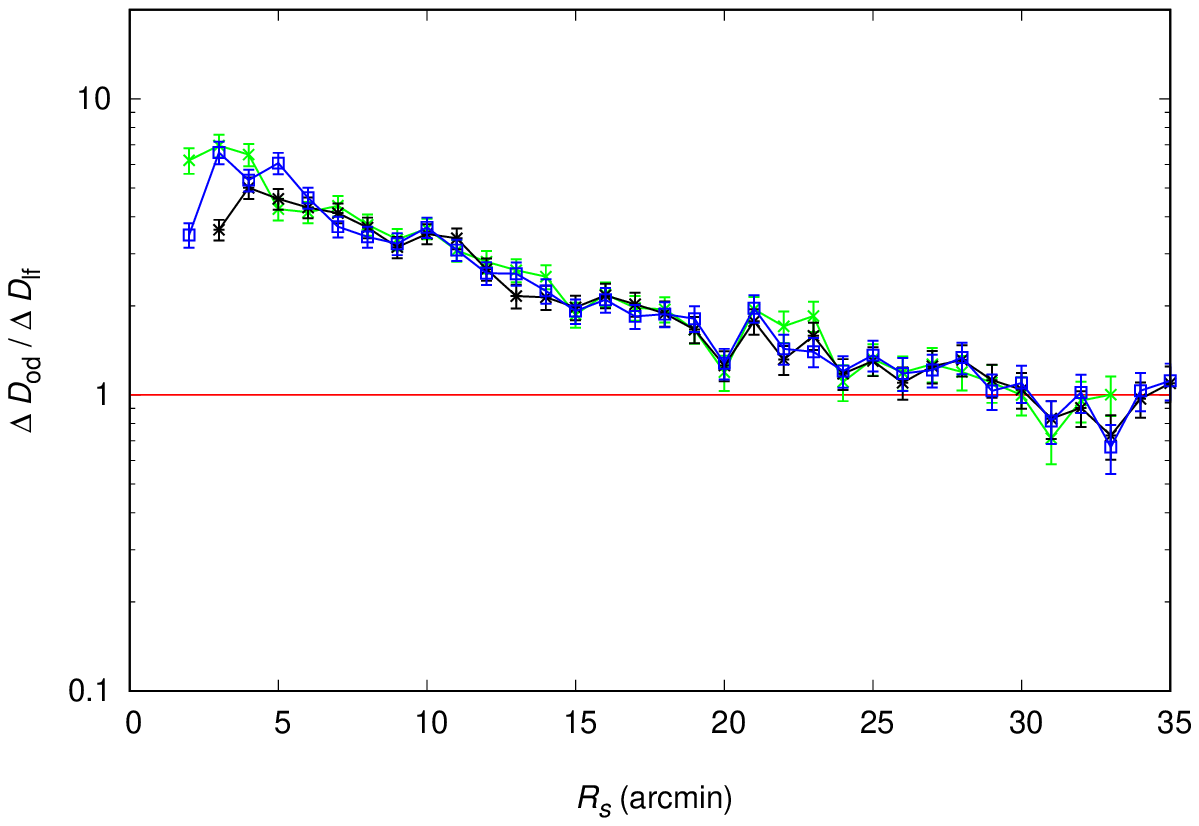}
\includegraphics[width=\columnwidth]{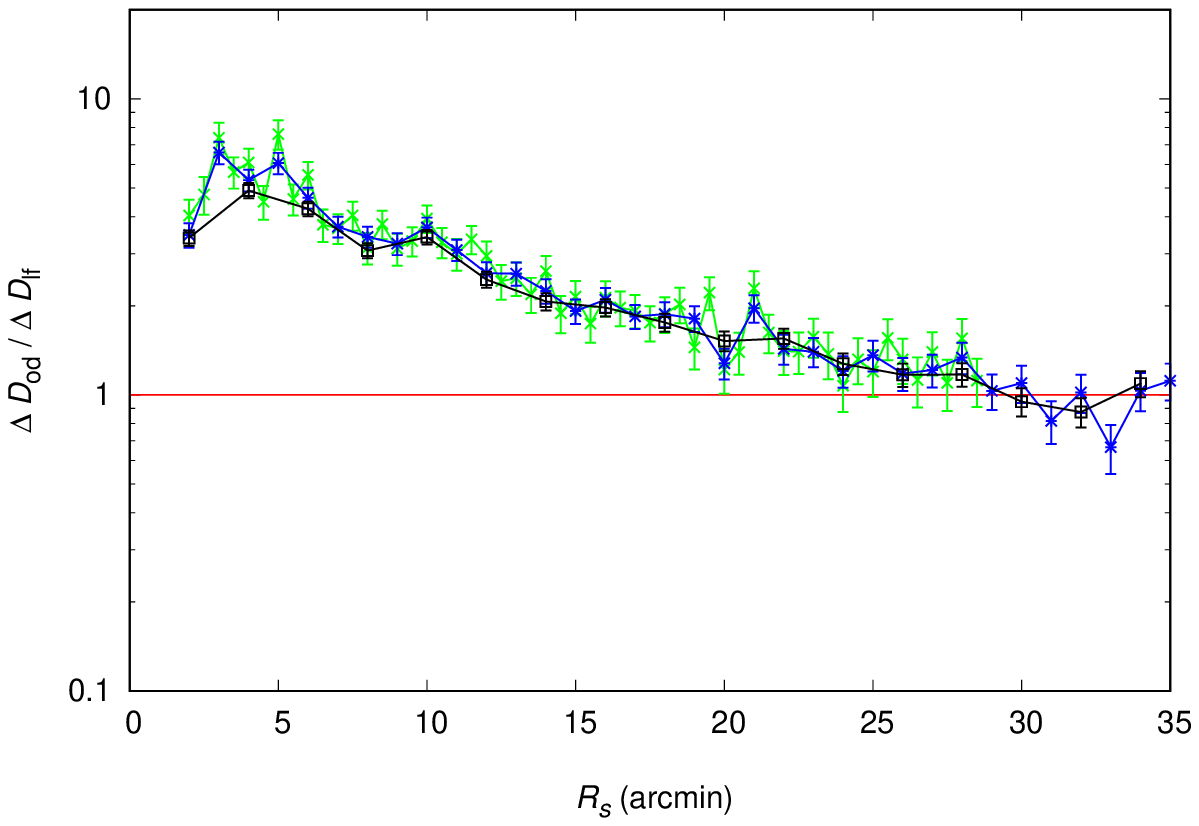}
\caption{Ratio of density variations ($\Delta D_{od} / \Delta 
D_{lf}$) of the open cluster NGC~2437 for different values of 
the free parameters. Left panel: for three values of the minimum 
number of allowed data points to estimate star densities in a given 
region: $N_{min}=50$ in green, $N_{min}=100$ in blue and $N_{min}=200$ 
in black. Right panel: for three values of the step in the sampling 
radius: $\delta R_{s}=0.5$~arcmin in green, $\delta R_{s}=1.0$ in 
blue and $\delta R_{s}=2.0$ in black. Uncertainties associated to 
$D_{lf}$ (grey regions in previous figures) are not shown here for 
clarity.}
\label{fig_param}
\end{figure*}

$N_{min}$ also determines, depending on the
projected cluster and field star densities, the
minimum starting value for $R_s$ (the value corresponding to
a sampling large enough). This means that if the cluster radius 
is smaller than the minimum $R_s$ then the algorithm will not find
it. In these cases $N_{min}$ should be decreased. After all the 
tests performed on simulated and real cluster, we have seen that 
when $N_{min} \lesssim 50$ the density estimations tend to be rather 
noisy and, therefore, we choose $N_{min}=100$ as the default value 
used in this work. In any case, final cluster radius values are not 
substantially affected by the exact value of this parameter. Left 
panel in Figure~\ref{fig_param} compares the $\Delta D_{od} / \Delta 
D_{lf}$ versus $R_s$ plots of the open cluster NGC~2437 for three 
different $N_{min}$ values.
The curves are practically the same. For the case $N_{min}=50$ 
(green line in Figure~\ref{fig_param}) the obtained cluster radius 
is $R_c=19-29$~arcmin, very similar to the range $R_c=22-29$~arcmin 
obtained for the rest of cases.

\subsection{Parameter $\delta R_s$}

The step size for increasing the sampling ($\delta R_s$)
is also a free parameter of the algorithm. 
A relatively small value for this parameter does not
necessarily mean a higher precision of the final $R_c$ value 
because, in this case, small changes in the radius imply small 
increments in the number of new sampled stars and, therefore, 
higher uncertainties and fluctuations in density change estimations. 
On the contrary, relatively high $\delta R_s$ values imply better 
density change estimations and produce smoother curves but at the 
expense of a smaller precision in the obtained $R_c$ value. Small 
or high $\delta R_s$ values are relative terms because they depend 
on the projected spatial density of stars and on the actual cluster 
radius value (that we cannot know a priori). All the tests performed 
with the data we are working with (the used list of clusters with 
proper motions from {\it Gaia} DR2) suggested that
$\delta R_s \sim 1$~arcmin is a good compromise
between both extremes and it is chosen as the starting 
value for the calculation. However, in order to ensure that solution 
convergence is achieved, the algorithm increases or decreases this 
initial value depending on the range of $R_s$ values to be explored 
and/or the average density of stars. As with the parameter $N_{min}$, 
the algorithm behaviour is not greatly affected by the exact value of 
the step in $R_s$. Right panel in Figure~\ref{fig_param} compares the 
results for the cluster NGC~2437 using three different values of this 
parameter. The algorithm yields noisier or smoother curves but that 
broadly follow the same pattern as the reference value ($\delta R_s=1$) 
although, obviously, the exact final $R_c$ estimation may differ slightly. 
In this case we get $25.5 \pm 3.5$ (for $\delta R_s=1$~arcmin), $26.0 \pm 
4.0$ ($\delta R_s=2$~arcmin) and $22.0 \pm 2..0$ ($\delta R_s=0.5$~arcmin), 
all compatible inside the error bars.

\subsection{Other parameters and conditions}

There may be significant differences in proper motion
errors between bright and faint sources. However, applying a magnitude
cut or filtering by proper motion error does not necessarily improve
the results. The reason is that, on the one hand, the quality
of the used data becomes better but, on the other hand, the number
of data points decreases and this results in more fluctuations
in the $\Delta D_{od} / \Delta D_{lf}$--$R_s$ plot.
In any case, for sufficiently well-sampled OCs we have
checked that, although the exact shapes of the curves might
differ, the obtained cluster radius remains unaltered (within
the calculated uncertainties) when filtering by magnitude or
errors in proper motion are applied. This is because the point
defining the cluster radius ($R_s=R_c$, at which no more cluster
stars can be added if $R_s$ is increased) does not depend on how
clearly the overdensity is seen in the proper motion space, as
long as it is properly detected.

The exact location of the centre of the cluster is also a
relevant issue. We have used cluster positions given by
\citetalias{Dia02} in their catalogue but positions given
in other catalogues or actual cluster centres may differ.
By using both simulated and well-behaved real clusters,
we have verified that, as expected, the shift of the centre
of the sampling circle relative to the cluster centre results
in an equivalent increase in the obtained cluster radius.
Then, if the exact cluster centre is unknown, the $R_c$
value obtained with the method proposed in this work should
be seen as an upper limit to the actual cluster radius.
Generally speaking, we expect this effect to be smaller
than the final radius uncertainty. Radius errors for
our 401 valid solutions distribute with a mean value
of $\sim 1.12$ arcmin (standard deviation $\sim 1.01$
arcmin), whereas angular distances (for the same 401 OCs)
between centres reported by \citetalias{Dia02} (also used
by \citetalias{Sam17}) and \citetalias{Kha13} distribute 
with a mean of $\sim 0.90$ arcmin (standard deviation
$\sim 0.67$ arcmin).

\bsp
\label{lastpage}
\end{document}